\documentclass{aa}         
\usepackage{times,graphics}

\usepackage{epsfig} 

\begin{document}
   
   \title{Understanding B-type Supergiants in the Low Metallicity Environment of the SMC.}
   
   \author{C. Trundle$^{1,2}$, D.J. Lennon$^{1}$, J. Puls$^{3}$, P.L. Dufton$^{2}$} 
   
   \offprints{C.Trundle at ct@ing.iac.es}

   \authorrunning{C.Trundle et al.}
   
   \titlerunning{SMC B-type Supergiants.}
   
   \institute{The Isaac Newton Group of Telescopes,
	      Apartado de Correos 321, E-38700,
	      Santa Cruz de La Palma, Canary Islands, Spain
		\and 
	      The Department of Pure and Applied Physics,
              The Queen's University of Belfast,
	      Belfast BT7 1NN, Northern Ireland
	      	\and
	      Universit\"{a}ts-Sternwarte M\"{u}nchen, 
	      Scheinstr. 1, D-81679, 
	      Germany. 
             }

   \date{}

   \abstract{Spectroscopic analyses of 7 SMC B-type supergiants and 1 giant have been
undertaken using high resolution optical data obtained on the  {\sc vlt} with {\sc
uves}. {\sc fastwind}, a non-LTE, spherical, line-blanketed model atmosphere
code was used to derive atmospheric and wind parameters of these stars as well
as their absolute abundances. Mass-loss rates, derived from H$_{\alpha}$
profiles, are in poor agreement with metallicity dependent theoretical
predictions. Indeed the wind-momenta of the SMC stars appear to be in good
agreement with the wind-momentum luminosity relationship (WLR) of Galactic
B-type stars, a puzzling result given that line-driven wind theory predicts a
metallicity dependence. However the galactic stars were analysed using
unblanketed model atmospheres which may mask any dependence on metallicity. A mean
nitrogen enhancement of a factor of 14 is observed in the supergiants whilst
only an enrichment of a factor of 4 is present in the giant, AV216. Similar
excesses in nitrogen are observed in O-type dwarfs and supergiants in the same
mass range, suggesting that the additional nitrogen is produced while the stars
are still on the main-sequence. These nitrogen enrichments can be reproduced by
current stellar evolution models, which include rotationally induced mixing,
only if large initial rotational velocities of 300 kms$^{-1}$ are invoked. Such
large rotational velocities appear to be inconsistent with observed $v \sin i$
distributions for O-type stars and B-type supergiants. Hence it is suggested
that the currently available stellar evolution models require more efficient
mixing for lower rotational velocities.} 

   \maketitle 

%

\section{Introduction} 
\label{intro} 

Due to its low metal content and close proximity the Small
Magellanic Cloud (SMC) is a key laboratory in which to study
massive stars.  Its metallicity ($Z$), or more accurately the
abundance of its iron peak and alpha-processed elements, is
approximately one fifth solar, making it an excellent template
for studies of metal poor starbursts and high redshift
star-forming  galaxies (Leitherer et al. 2001, Pettini et al.
2000).  Its proximity, at a distance modulus of 18.9 (Harries,
Hilditch, \& Howarth 2003), makes high resolution studies
feasible with 4--8 m telescopes and since the extinction is low,
E(B-V)$\sim$0.07, similar studies are also feasible in the
important ultraviolet regime. 

This combination of circumstances has led to a number of detailed
studies of O-type stars in the SMC at optical and UV wavelengths
(Walborn et al. 1995; Puls et al. 1996; Haser et al. 1998).   Even
more recently we have seen the publication of work related to a high
resolution UV study of a sample of O-stars in the SMC (Walborn et
al. 2000; Bouret et al. 2003; Hillier et al. 2003).  In the present
paper, and a companion paper (Evans et al. 2004; hereafter EVANS04),
we extend this work into the domain of the B-type stars in the SMC. 
There are important reasons for this extension, among them the fact
that the B-type stars are important contributors to the UV spectrum
of continuous starforming and intermediate age (20-50 Myr) stellar
populations (deMello, Leitherer, \& Heckmann 2000).  In fact the UV
spectra of the stars discussed in the present paper have been used
to constrain the age of the metal poor super star cluster 
NGC1705-1 (Vazquez et al. 2003). However what is more relevant to
the present paper is that B-type supergiants in the SMC provide
important information on mass-loss and stellar evolution in a low
metallicity environment.

The surface compositions of B-type supergiants are a key test of stellar
evolution models since mixing and mass-loss processes may alter their
surface CNO abundances, especially that of nitrogen.  The SMC is an ideal
laboratory in this respect since its pristine nitrogen abundance is about
1/30th to 1/50th of solar and enhancements are correspondingly easier to
detect. Enhanced nitrogen abundances have already been observed in
Galactic and Magellanic Cloud B-type supergiants indicating some mixing of
CNO-cycled material to the stellar surface (Gies \& Lambert 1992; Lennon
et al. 1991, 1996, 1997, 2003; Fitzpatrick \& Bohannan 1993; McErlean,
Lennon, \& Dufton 1999; Dufton et al. 2000). Moreover, Lyubimkov
(1991) and later Gies \& Lambert (1992) found that Galactic main-sequence
B-type stars also show such N enhancements in their spectra.  Whether the
mechanism which causes this mixing of core processed material to the
photosphere is the presence of a blue-loop evolutionary phase (Schaller et
al. 1992) or due to stellar rotation (Denissenkov 1994; Talon 1997) is
still uncertain. At present much effort is being put into the development
of stellar evolution models which include rotationally induced mixing
(Heger \& Langer 2000; Maeder \& Meynet 2000, 2001).  Such models have had
success in predicting the correct dependence of the blue to red
supergiant (B/R) ratio on metallicity, previously a problem for models
without rotation. This ratio is an important factor in stellar and
galactic evolution models as it represents the relative lifetimes of hot
and cool core helium burning phases. Eggenberger, Meynet, \& Maeder (2002)
showed that evolution models including rotationally induced mixing (Maeder
\& Meynet 2000) seem to reproduce the observed B/R ratio at Galactic, LMC
and SMC metallicities, as suggested by Langer \& Maeder (1995).  Lennon,
Dufton, \& Crowley (2003; hereafter LEN03) investigated surface abundances
of a sample of lower luminosity B-type giants in the SMC, and discussed
these in the context of evolution with rotation. Whether these
evolutionary models can reproduce the enhanced nitrogen abundances and
rotational velocities observed in supergiants/giants in a low metallicity
regime will be explored in this paper.

Blue supergiant stars, by virtue of their low bolometric correction and high
intrinsic luminosity, are the most accessible stars to study at visible
wavelengths.  The first chemical compositions of individual stars in the
Magellanic Clouds were determined from  B- to F-type supergiants by Przybylski
(1968, 1971), Wolf (1972, 1973) and Osmer (1972, 1973), but with mixed
results.  Osmer (1973) concluded, using simple curve of growth arguments, that
since the metal lines in a B1.5 and B2.5 supergiant were approximately a factor
of two lower than their Milky Way counterparts then their metallicities must be
at least a factor of four below solar. This is remarkably close to the now
commonly accepted value of one-fifth solar and its accuracy is probably due in
part to the  care with which galactic standards were chosen as comparison
stars.  Subsequently blue supergiants have provided important insights into the
current chemical makeup of many galaxies in the Local Group and beyond
(Monteverde et al. 1996, 1997, 1998, 2000; Venn et al. 1995, 1999, 2000, 2001,
2003a, 2003b; McErlean, Lennon, \& Dufton 1999; Dufton et al. 2000; Smartt et
al. 2001; Bresolin et al. 2001, 2002a, 2002b; Trundle et al. 2002; Urbaneja et
al. 2003) using somewhat more sophisticated modeling techniques.  Lennon et al.
(1991) used plane-parallel non-LTE methods to analyse a B-type supergiant in
each Cloud, this being followed by a more comprehensive non-LTE study of 48
B-type supergiants in the SMC by Dufton et al. (2000).  Nevertheless as
discussed by  McErlean, Lennon, \& Dufton (1998) the effect of microturbulence
and the neglect of the stellar wind in this plane-parallel approach are
significant sources of uncertainties in their results. With the recent
development of stellar atmosphere codes which treat in a unified manner both
the photosphere and wind of massive stars, it is now possible to include the
effects of the stellar wind (Santolaya-Rey, Puls, \& Herrero 1997; Herrero,
Puls, \& Najarro 2002; Repolust, Puls, \& Herrero 2003 - {\sc fastwind};
Hillier \& Miller 1998 - {\sc cmfgen}; Pauldrach, Hoffmann, \& Lennon 2001 -
wm-basic). We use {\sc fastwind} in the present paper which in addition to
determining the chemical composition of our sample, enables us to derive
estimates for their mass-loss rates and wind momenta.

Kudritzki et al. (1995) introduced the Wind-Momentum Luminosity Relationship
(WLR) as a method of determining distances out to the Virgo and Fornax Clusters
through the spectral analysis of hot, massive stars.  Line-driven wind theory
implies that photon momentum transfer through metal line absorption is driving
the stellar wind. As the WLR is based on line-driven wind theory, it follows
that this relationship should be dependent on metallicity.  Studies of O-type
supergiants in the SMC and LMC have confirmed this metallicity dependence. 
Additionally, analyses of Galactic early-type supergiants have shown some
dependence on spectral type (Puls et al. 1996; Kudritzki et al. 1999).  More
recent analyses of Bouret et al. (2003) and Repolust, Puls, \& Herrero (2003)
show a distinct difference in the wind momenta of O-type supergiants to that of
giants and dwarfs in both the Galaxy and Magellanic Clouds. In this paper we
will compare the WLR of our sample with that of stars in the Galaxy and with
the predictions of theory.

%

\section{Observations and Data Reduction}

\begin{table*}
\caption[]{ 
Observational details. Identification numbers are from  Azzopardi \&
Vigneau (1982; AV\#) and Sanduleak (1968; Sk\#). Spectral types
adopted from Lennon (1997). Absolute Magnitudes (M$_{v}$) are
calculated using V and (B-V) magnitudes from Garmany, Conti
\& Massey (1987;$^{1}$) and  Massey (2002;$^{2}$) and (B-V)$_{0}$ values from
Fitzpatrick \& Garmany (1990). The adopted distance modulus is 18.9
(Harries, Hilditch, \& Howarth 2003). The average S/N ratios for the
{\sc vlt/uves} data are given, but the actual S/N varied across the
orders. Heliocentric corrected radial velocities are presented for
both the {\sc vlt/uves} and {\sc ntt/emmi} datasets. Note there is
no significant variation in radial velocities from the two datasets
and thus no obvious indication of binary stars amongst the sample. In
addition we note that the quoted $v \sin i $ values are upper limits
and are dominated by macroturbulence as discussed in Sect. 2.
}
\begin{flushleft}
\centering
\begin{tabular}{lllcccccccc} \hline \hline
STAR & Alternative & Spectral & V & B-V & M$_{v}$ & \multicolumn{2}{c}{S/N} & \multicolumn{2}{c}{v$_{lsr}$}& $v \sin i $  \\
     & ID          & Type     &   &     &         &  B   & R         
           &	{\sc uves} & {\sc emmi}   &(kms$^{-1}$)\\ 
\hline   
\\            
AV215 & Sk76  & BN0 {\sc I}a  & 12.69 $^{2}$& -0.09 & -6.61 & 120 & 110 & 154 $\pm$ 10 &  159 $\pm$ 15 & 91
\\
AV104 &       & B0.5 {\sc I}a & 13.17 $^{2}$& -0.16 & -5.82 & 150 & 120 & 163 $\pm$ 12 &  165 $\pm$ 28 & 80
\\
AV216 &       & B1 {\sc III}  & 14.22 $^{2}$& -0.13 & -5.08 & 120 & 100 & 203 $\pm$ 14 &             & 77
\\
      & Sk191 & B1.5 {\sc I}a & 11.86 $^{1}$& -0.04 & -7.41 & 170 & 140 & 130 $\pm$ 24 &  134 $\pm$ 17 & 96
\\
AV210 & Sk73  & B1.5 {\sc I}a & 12.60 $^{2}$& -0.02 & -6.73 & 180 & 140 & 173 $\pm$  8 &  192 $\pm$ 12 & 65
\\   
AV18  & Sk13  & B2 {\sc I}a   & 12.46 $^{1}$&  0.03 & -7.00 & 170 & 120 & 148 $\pm$ 10 &  127 $\pm$ 13 & 49
\\
AV362 & Sk114 & B3 {\sc I}a   & 11.36 $^{1}$& -0.03 & -7.82 & 190 & 100 & 208 $\pm$ 14 &  214 $\pm$ 10 & 51
\\
AV22  & Sk15  & B5 {\sc I}a   & 12.25 $^{1}$& -0.10 & -6.62 & 140 & 120 & 139 $\pm$ 12 &  109 $\pm$ 16 & 46
\\
\hline             
\end{tabular}
\end{flushleft}
\label{obsdata} 
\end{table*}     

High resolution echelle spectra were obtained for eight SMC B-type stars
covering the spectral range of B0 - B5, using the {\sc uves} spectrograph on
the {\sc vlt}. The data were taken during twilight on a 3 night run in November
2001, see the observational details in Table~\ref{obsdata}. {\sc uves} is a two
arm crossdispersed echelle spectrograph, the red arm of which contains a mosaic
of two chips.  EEV CCD-44 2 k*4 k chips are used in both arms along with a 
MIT-LL CCID-20 CCD chip in the red arm which reduces the fringing beyond 7000
\AA .  The useful wavelength ranges obtained from the three chips were 3750 -
5000 \AA, 5900 - 7700 \AA\ \& 7750 - 9600 \AA. These CCD chips have a pixel
size of 15 $\mu$m and a slit width of 1.5 arcsec which was used to give a
resolving power of approximately 20,000. The spectra were later binned to a
dispersion resolution of 0.2 \AA\ per pixel. 

The 2-dimensional CCD frames were transformed to multiple order spectra using
the {\sc uves} pipeline software. This was done in an interactive mode through
the {\sc echelle/uves} context of  {\sc midas}.  Th-Ar arc lamp exposures were
used to define echelle order positions and wavelength calibrate the spectra. A
constant bias level and median background were subtracted from the flat-fields
which were then extracted in pixel-order space. The stellar spectra were
reduced using both average and optimal extraction methods. In the blue region
our analysis utilised the optimally extracted spectra, however due to the
magnitude of the fringing in the red region the averaged extracted data proved
to be more useful. The final merged spectra and individual orders were then
processed using an automated {\sc idl} routine to filter out the cosmic rays. 

\begin{table*}
\caption[]{
Estimated equivalent widths (m\AA) for SMC B-type supergiants.
Uncertainties range from 10 \% for the best observed features to 20
\% for the weaker features. 
}
\begin{flushleft}
\centering
\begin{tabular}{lcccccccc} \hline \hline
Line & AV215  & AV104 & AV216 & Sk191 & AV210 & AV18 & AV362 & AV22
\\
\hline
\\
C {\sc ii}  4267 & 25 & 52: & 49 & 48  & 83  & 81  & 53  & 55
\\
C {\sc ii}  6578 &    &     &    &     & 105 & 154 & 120 & 72
\\
C {\sc ii}  6580 &    &     &    &     & 116 & 119 & 70  & 77
\\
N {\sc ii}  3995 & 68 & 49  & 40 & 137 & 173 & 169 & 95  & 81
\\
N {\sc ii}  4447 &    &     &    & 47  & 69  & 47  & 39  & 26
\\
N {\sc ii}  4601 &    &     &    & 50  & 59  & 49  &     & 
\\
N {\sc ii}  4607 &    &     &    & 29  & 61  & 46  &     &
\\
N {\sc ii}  4614 &    &     &    &     & 42  & 22  &     &
\\
N {\sc ii}  4621 & 31 &     &    & 33  & 51  & 25  &     & 
\\
N {\sc ii}  4630 & 73 & 48  &    & 95  & 158 & 103 & 53  & 47
\\
O {\sc ii}  4072 & 57 & 88  & 87 & 96  & 86  & 69  &     & 
\\
O {\sc ii}  4076 & 44 & 121 & 87 & 141 & 106 & 88  &     & 
\\
O {\sc ii}  4317 &    & 40  &    & 95  & 72  & 61  &     & 
\\
O {\sc ii}  4320 &    & 36  &    & 87  & 87  & 63  &     & 
\\
O {\sc ii}  4367 & 38 & 49  &    & 92  & 88  & 61  &     & 
\\ 
O {\sc ii}  4415 &    & 121 & 86 & 166 & 132 & 109 &     & 
\\
O {\sc ii}  4417 &    & 31  & 70 & 71  & 81  & 4066  &     & 
\\
O {\sc ii}  4591 & 54 & 67  & 29 & 86  & 58  & 40  &     & 
\\
O {\sc ii}  4596 & 58 & 50  & 43 & 62  & 43  & 33  &     & 
\\
O {\sc ii}  4638 &    & 40  &    & 141 & 81  & 169 &     & 
\\ 
O {\sc ii}  4641 &    & 137 &    & 250 & 92  & 93  &     & 
\\
O {\sc ii}  4662 & 55 & 69  & 50 & 151 & 207 & 167 &     & 
\\
Mg {\sc ii}  4481& 80: & 65  & 46 & 75  & 114 & 93 & 182 & 217
\\
Si {\sc ii} 4128&    &     &    &     &     & 30: & 73  & 94
\\
Si {\sc ii} 4131&    &     &    &     &     & 70: & 79  & 96
\\
Si {\sc iii} 4553& 168& 161 & 94 & 266 & 264 & 199 & 56  & 58
\\
Si {\sc iii} 4568& 129& 120 & 78 & 206 & 211 & 174 & 47  & 45
\\
Si {\sc iii} 4575&  64 &  54 & 40 & 108 & 113 & 95  & 21 &
\\
Si {\sc iv}  4116& 212& 111 & 60 & 47 & 34 &   &   & 
\\
\hline
\end{tabular}
\end{flushleft}
\label{ew}
\end{table*}

Normalisations, corrections for radial velocity shifts and
equivalent width (EW) measurements were carried out in {\sc dipso},
the spectrum analysis procedure of {\sc starlink}. The measured
equivalent widths for the lines used in the abundance analysis are
given in Table~\ref{ew}.  Although the {\sc reduce/uves} routine
normalised and merged the echelle orders, these merged spectra were
checked manually.  Low order polynomials were fitted to the
continuum of each order for normalisation, the orders were then
merged using appropriate weightings. In most cases, the
automatically and manually reduced data were in good agreement.
Nevertheless in measuring the equivalent widths of lines in the
overlapping regions of different orders and in analysing the Balmer
lines, more weight was given to the manually reduced spectra.  

The metal and diffuse helium absorption lines were fit using a
non-linear least squares technique with a Gaussian profile. For
isolated, well observed lines the full-width-half-maximum (FWHM) and
position were allowed to vary.  Blended lines such as the C {\sc ii}
4267 \AA\ and Mg {\sc ii} 4481 \AA\ were treated as multiple
Gaussians with fixed wavelength separations and adopting the mean
FWHMs.  For well observed lines error estimates of 10 \% are
appropriate, however blended and weak lines have errors of 20 \% or
more.  The signal-to-noise ratio (S/N) obtained was measured
throughout the spectral range of each arm in regions of continuum
with no observed spectral lines. In the blue region S/N is typically
150 or higher and in the red all spectra have a S/N greater than
100. Radial velocity shifts were measured from the positions of the
well observed lines and are presented in Table~\ref{obsdata} along
with the mean S/N ratios.

Additionally, medium resolution spectra of the seven supergiants were
previously obtained using the NTT telescope in remote observing mode with the
ESO Multi Mode Instrument ({\sc emmi}). The reduction methods of this dataset
were discussed in Lennon (1997) and all equivalent widths were remeasured here
using the technique described above. Each object was observed for between 10 -
15 mins. Depending on its visual magnitude and the weather conditions at the
time, these exposure times gave signal-to-noise ratios greater than 60 in the
red and 90 in the blue. The equivalent widths measured from this dataset agree
with the {\sc uves} dataset to within 5 - 20 \% depending on the quality of the
spectra. Comparison of radial velocity measurements from the {\sc emmi} and
{\sc uves} datasets show no signs of binaries amongst the stellar selection
(see Table~\ref{obsdata}). However from STIS/HST target acquisition images,
EVANS04 noticed a companion to AV216, the faintness of which suggests that it
is a main-sequence star. Indeed some asymmetry was observed in the H$_{\alpha}$
and He {\sc i} 4471 \AA\ profiles (see Figs.~\ref{ha} \& \ref{blue2}). The
additional dataset was mainly used for a check on our {\sc uves} data. However
due to some nebular contamination of the H$_{\alpha}$ profiles of AV362 \&
AV22, the {\sc emmi} dataset was used to derive the mass-loss rate of these two
stars. There was also some nebular contribution observed in the H$_{\alpha}$
profile of Sk191, and we have taken this into account in fitting the data.

The projected rotational velocity for each star, $v \sin i$, was determined
from a selection of seven unblended metal lines by fitting theoretical profiles
to the observed spectra. The theoretical profiles were convolved with a
Gaussian profile to account for the instrumental broadening.  The profile was
then convolved with a rotational line profile function, this has been described
in detail by Rucinski (1990) and more recently by Gray (1992). Howarth et al.
(1997) and Ryans et al. (2002) found that the broadening of spectral features
in B-type supergiants was dominated by macroturbulence. Indeed our projected
rotational velocity estimates listed in Table~\ref{obsdata} show a similar
behaviour. Ryans et al. (2002) also estimated that the typical projected
rotational velocity was of the order 10-20 kms$^{-1}$. Hence the values for $v
\sin i$ tabulated in Table~\ref{obsdata} should be considered as indicative of
the width of the spectral features and as upper limits on the projected
rotational velocities.

%
\section{Model Atmospheres}
'Unified model atmosphere' codes such as those mentioned in the
introduction have an advantage over plane-parallel codes (viz. {\sc
tlusty} - Hubeny \& Lanz 1995). This is due to  treating the
atmosphere in a consistent manner by including a smooth transition,
from a pseudo-hydrostatic photosphere to an outer expanding
atmosphere - 'the wind'. We have opted to use {\sc fastwind} (Fast
Analysis of STellar atmospheres with WINDs), a spherically symmetric
non-LTE code first introduced by Santolaya-Rey, Puls, \& Herrero
(1997). {\sc fastwind} has been developed to include an approximated
treatment of metal line -blocking and -blanketing (Herrero, Puls, \&
Najarro 2002; Repolust, Puls, \& Herrero 2003).

\begin{table*}
\caption[]{
Derived atmospheric parameters for SMC B-type supergiants. The
microturbulence from both the Si {\sc iii} \& O {\sc ii} lines
are presented, however the silicon microturbulence was adopted
in our analysis. A microturbulence of 10 kms$^{-1}$ was adopted
for AV22, due to the difficulties in constraining $\xi$ from
the metal lines in its' spectrum. M$_{evol}$ are estimated from the 
stellar evolution tracks of Maeder \& Meynet (2001), see discussion
in Sect. 6.2. The errors quoted are the
typical random errors as discussed in Sect. 4. 
}  
\begin{flushleft}
\centering
\begin{tabular}{llcccccccc} \hline \hline
Star & Spectral &T$_{\rm eff}$ & $\log$ g & R$_{\star}$&M$_{spec}$&M$_{evol}$ & $\log(\frac{L_{\star}}{L_{\odot}})$ & $\xi$$_{\rm Si}$ & $\xi$$_{\rm O}$   \\
     & Type       &  (kK)         &   (cgs) & (R$_{\odot}$)    &(M$_{\odot}$)&(M$_{\odot}$)&                              & \multicolumn{2}{c}{ (kms$^{-1}$)}    \\                                        
\hline   
\\            
AV215 & BN0 {\sc I}a  & 27.0 $\pm$ 1.0 $^{1}$ & 2.90 $\pm$ 0.10 & 30 & 26 & 39 & 5.63 & 12 & 17 
\\
AV104 & B0.5 {\sc I}a & 27.5 $\pm$ 1.0 $^{1}$ & 3.10 $\pm$ 0.10 & 20 & 19 & 26 & 5.31 & 11 & 20 
\\
AV216 & B1 {\sc III}  & 26.0 $\pm$ 1.5 $^{2}$ & 3.60 $\pm$ 0.20 & 16 & 36 & 20 & 5.00 & 0  & 15  
\\
Sk191 & B1.5 {\sc I}a & 22.5 $\pm$ 1.5 $^{2}$ & 2.55 $\pm$ 0.15 & 51 & 33 & 41 & 5.77 & 13 & 20 
\\
AV210 & B1.5 {\sc I}a & 20.5 $\pm$ 1.5 $^{2}$ & 2.40 $\pm$ 0.15 & 40 & 15 & 27 & 5.41 & 12 & 18  
\\   
AV18  & B2 {\sc I}a   & 19.0 $\pm$ 2.0 $^{3}$ & 2.30 $\pm$ 0.20 & 49 & 17 & 28 & 5.44 & 9  & 16  
\\
AV362 & B3 {\sc I}a   & 14.0 $\pm$ 1.5 $^{4}$ & 1.70 $\pm$ 0.15 & 96 & 17 & 30 & 5.50 & 10	  
\\
AV22  & B5 {\sc I}a   & 14.5 $\pm$ 1.5 $^{4}$ & 1.90 $\pm$ 0.15 & 53 &  8 & 19 & 5.04 & 10$^{*}$ 
\\
\hline             
\multicolumn{10}{l}{\footnotesize{T$_{\rm eff}$
determined from : $^{1}$ Si {\sc iii}/Si {\sc iv} ionisation balance \& He {\sc ii} fits ; 
 }}\\
\multicolumn{10}{l}{\footnotesize{$^{2}$ Si {\sc iii}/Si {\sc iv} ionisation balance ; $^{3}$ Fits to Si {\sc iii}
lines ; $^{4}$ Si {\sc ii}/Si {\sc iii} ionisation balance.}}\\
\end{tabular}
\end{flushleft}
\label{smcpar1} 
\end{table*}  

The main advantage of {\sc fastwind} over more sophisticated
codes such as {\sc cmfgen} (Hillier \& Miller 1998) is the short
computational time it takes to calculate models for a significant
parameter space.  Fast computational time allows for quantitative
spectroscopic analyses of large samples of massive stars whilst
still accounting for the stellar wind. This is achieved by
parameterising the temperature structure using the non-LTE Hopf
function described by Santolaya-Rey, Puls, \& Herrero (1997) which
includes the effects of sphericity and mass-loss: 

\begin{equation}
q\prime_{N}(\tau\prime_{R}) = \frac{\tau\prime_{R}}{\tau_{R}}q_{N}(\tau_{R})
\end{equation}

where $\tau\prime_{R}$ and q$\prime_{N}$ are the spherical
generalisation of the optical depth, $\tau_{R}$, and the plane
parallel Hopf function, q$_{N}(\tau_{R})$. q$_{N}(\tau_{R})$ and
$\tau\prime_{R}$ are defined by

\begin{equation}
q_{N}(\tau_{R}) \approx q_{\infty} + (q_{o} -q_{\infty})exp(-\gamma\tau_{R}) 
\end{equation}

and

\begin{equation}
d\tau\prime_{R} = \chi_{R}(r)(\frac{R_{\star}}{r})^{2}dr
\end{equation}

where q$_{\infty}$, q$_{o}$ and $\gamma$ are fit parameters to the run of the
Hopf function, $q_{N}$, as a function of Rosseland opacity, ($\chi_{R}$). For
the models discussed in this paper we have adopted Hopf parameters from the
parameterisation of non-LTE line-blanketed (including Iron) {\sc tlusty} models
calculated at Queen's University Belfast (Ryans, R.S.I.; private
communication). The {\sc tlusty} grid covered a temperature range of 15 kK to
33 kK and logarithmic surface gravities from 4.5 dex down to the Eddington
limit. The abundances in this grid were baseline SMC abundances (viz.
[Fe/H]=6.80 dex). The flux conservation reached by using the {\sc tlusty}
temperature stratification  was typically of the order of 1 - 2 \%, however for
stars with high mass-loss rates this was higher at $\sim$ 3 \%.

{\sc fastwind} has a similar philosophy to that of the line
formation code {\sc detail} (Butler \& Giddings 1985), i.e the code
is data driven. The calculated models for this work include atomic
data for hydrogen, helium, carbon, nitrogen, oxygen, magnesium  and
silicon. The atomic models and Stark broadening data implemented in
the line formation of hydrogen and helium are described in full in
Santolaya-Rey, Puls, \& Herrero (1997). In the case of silicon,
nitrogen, and oxygen, the metal ion populations and line profiles
were calculated using the atomic data of Becker \& Butler (1988,
1989, 1990). The atomic data used for C\,{\sc ii} are from Eber \&
Butler (1988), and the Mg\,{\sc ii} is from Mihalas (1972). In most
cases these datasets needed some modification to ensure  convergence
for the temperatures and gravities considered here,  however tests
showed that these had negligible impact on the resultant line
strengths. 

For Si\,{\sc ii} we initially used the silicon model atom as
described by Becker \& Butler (1990). This model essentially uses
the atomic levels adopted by Kamp (1978), but with updated atomic
data for bound-bound and bound-free transitions (see also Lennon et
al. 1986).  However comparisons of {\sc tlusty} and {\sc fastwind}
calculations in the low mass-loss rate limit (see below) showed
severe discrepancies for the Si\,{\sc ii} equivalent widths.  This
was traced to the simplified atomic model for this ion in the {\sc
fastwind} calculations.  We therefore replaced our 12-level 
Si\,{\sc ii} model ion with a more complete 34-level model ion. This
has important consequences for the determination of the effective
temperatures of B-type stars, in that for the mid to late B-type
stars (B2 - B9) the derived temperatures will be cooler using the new
Si {\sc ii} model atom. However it is beyond the scope of this paper
to include a detailed discussion of this and so it will be reported
in future work.

To check the validity of our results from {\sc fastwind}, we have
carried out preliminary tests with the plane-parallel code {\sc
tlusty} and unified model atmosphere code, {\sc cmfgen} (Hillier \&
Miller 1998). Choosing a star with a thin-wind, AV104, and a
thick-wind, AV215, we analysed these stars with {\sc tlusty} and
{\sc cmfgen}, respectively. No major discrepancies were found in
these analysis with any differences being within the estimated
errors. This is reassuring given the different atomic data and more
importantly the various physical assumptions each code adopts (viz.
degree of line-blanketing, treatment of wind, calculation of formal
solution etc).

Finally it should be noted that some of the projected rotational
velocities listed in Table~\ref{obsdata} are relatively large.
Therefore, an approach allowing for the variation of atmospheric
structure over the stellar surface (as used for example by Howarth \&
Smith 2001) might be warranted. However as discussed above, the actual
rotational velocities are likely to be considerably smaller and the
assumption of spatial homogeneity should be appropriate.

\section{Photospheric and Wind Analysis}

\begin{figure*}
\begin{center}
\epsfig{file=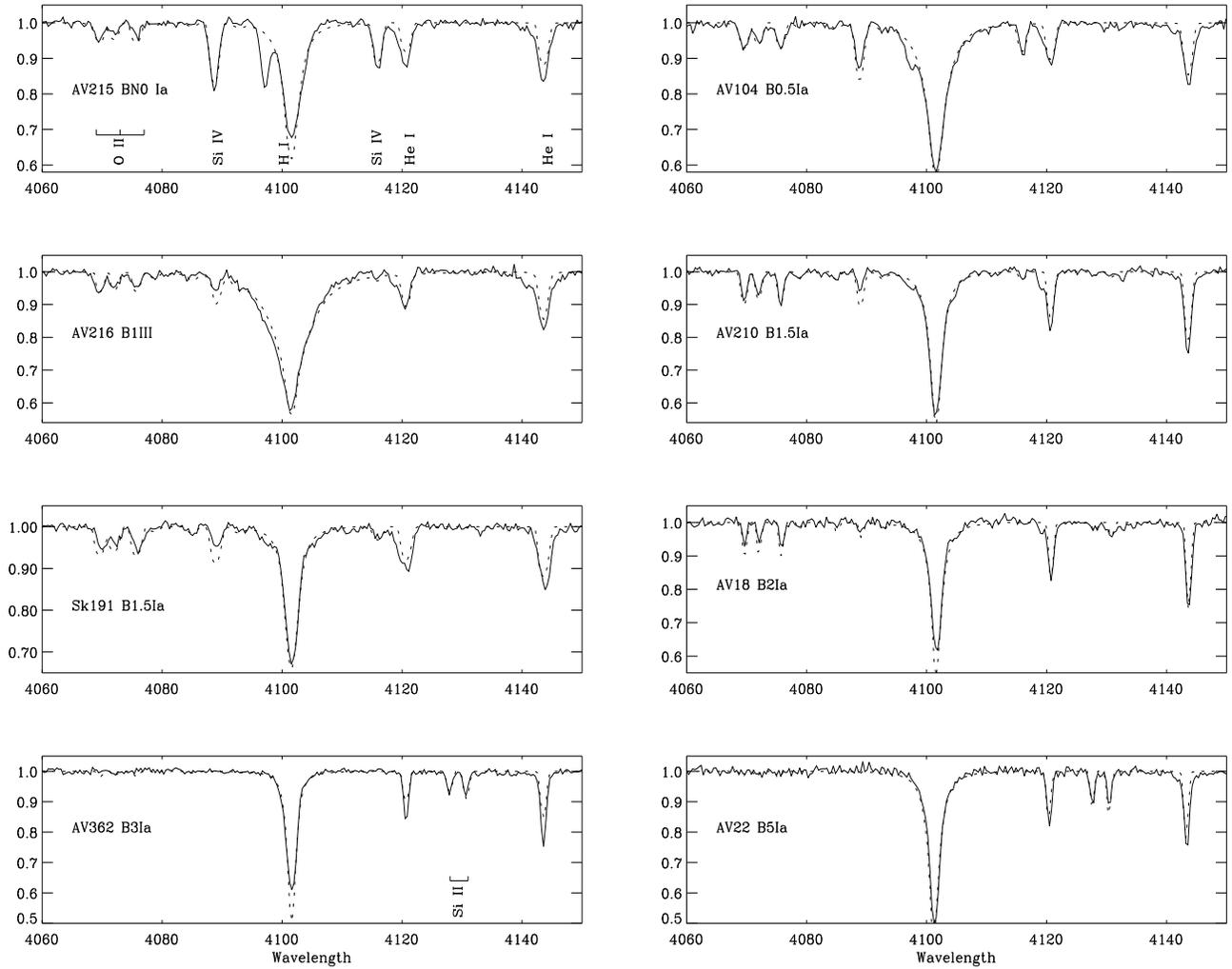, height=140mm, width=180mm, angle=0}
\caption[]
{Profile fits to spectral region 4060 - 4150 \AA. Observed
and theoretical data are shown as solid (-) and dotted
($\cdot\cdot\cdot$) lines, respectively. Note the fits to the
wings of the H$_{\delta}$ line which were used to fix $\log$ g and
the varying strengths of the Si {\sc ii \& iv} lines that were
the temperature indicators. Also note the absence of the O
{\sc ii} multiplet at 4070 \AA\ in the coolest stars AV362 \&
AV22. O {\sc ii} is not observed in the spectra of such cool
stars, preventing the determination of $\xi_{0}$.  
}
\label{blue1}
\end{center}
\end{figure*}


\subsection{Stellar parameters}

\begin{figure*}
\begin{center}
\epsfig{file=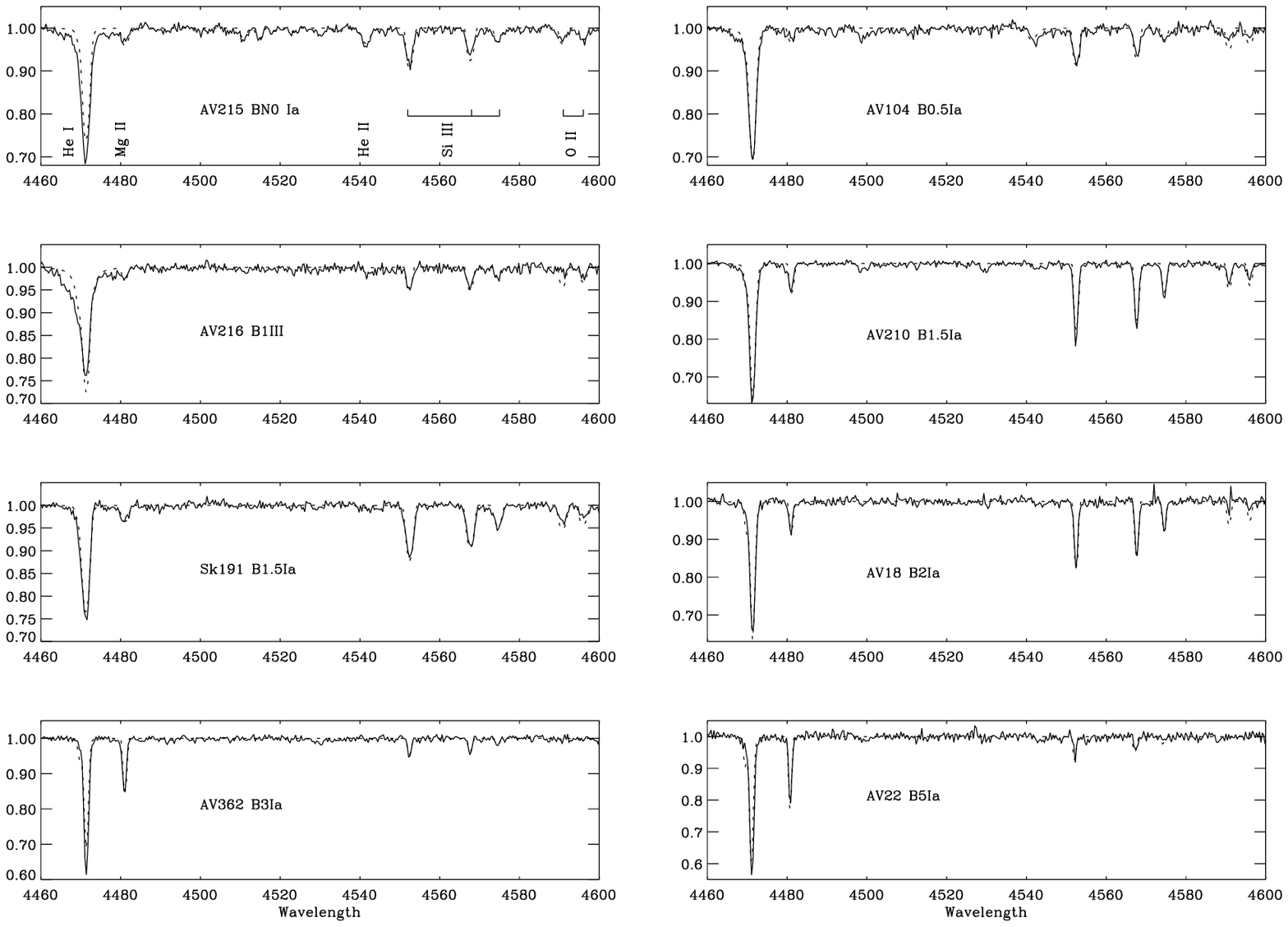, height=140mm, width=180mm, angle=0}
\caption[]
{
Profile fits to spectral region 4460 - 4600 \AA. Observed  and
theoretical data are shown as solid (-) and dotted ($\cdot\cdot\cdot$) lines,
respectively. Note the varying strengths of the Si {\sc iii} lines, which
were used  as temperature indicators. Also note the presence of the He {\sc ii}
line at 4542 \AA\ in AV215 and AV104, which was used as an additional
temperature diagnostic. The asymmetry in the He {\sc i} 4471 line of AV216
suggests some multiplicity. 
}
\label{blue2}
\end{center}
\end{figure*}

The atmospheric and wind parameters of our stellar sample are
presented in Tables~\ref{smcpar1} \&~\ref{smcpar2},
respectively. The methods for determining these parameters are
described in the following section; however one should note that
they are not derived independently but through an iterative
process.

In order to allow an initial assessment of the parameter space
appropriate for each of the SMC stars, a coarse grid of {\sc
fastwind} models were calculated to cover the spectral range B0 -
B5.  For each spectral bin, we adopted one effective temperature
and a range of surface gravities, spanning 0.5 dex. The
microturbulence adopted was 10 kms$^{-1}$ as appropriate for
B-type supergiants, if derived from the silicon spectrum (see
Sect. 4.1.3). The stellar radius was chosen based on the
effective temperature of the model as well as bolometric
corrections and colors taken from Fitzpatrick \& Garmany (1990).
The $\beta$-parameter was fixed at 1 and terminal velocities were
taken from Kudritzki \& Puls (2000), where a calibration of
terminal velocities with spectral type for Galactic stars is
presented. Five mass-loss rates for each spectral bin were
adopted, covering thin to thick winds.  To obtain initial
estimates of the parameters for each star, the appropriate set of
models for its spectral type was chosen and a comparison with the
observed data gave an approximate mass-loss rate and surface
gravity. Once initial values were defined, finer, smaller grids
were run which were appropriate to each individual star. 


\subsubsection{\it Effective temperature, T$_{\rm eff}$}

In early B-type stars, the silicon lines are the spectral
features most sensitive  to changes in effective temperature,
T$_{\rm eff}$ (see Figs.~\ref{blue1} \&~\ref{blue2} for the
variation of the  silicon ions line strength with spectral type).
The standard method of determining T$_{\rm eff}$ in B-type stars
is to search for an ionisation balance between consecutive stages
of silicon ionisation (Kilian et al. 1991). Although these lines
are sensitive to the silicon abundance, by considering the ratio of
the EW's for neighbouring ionisation stages we eliminate, to first
order, this dependence on abundance.  Initially T$_{\rm eff}$ was
constrained by fitting theoretical profiles to the Si lines, this was
then followed by a comparison between EW's of  theoretical and
observed profiles.  The error in this method due to observational
constraints and random errors is typically $\pm$ 1.5 kK, with a
few exceptions as discussed below.

For the hotter stars of our sample (i.e. B0 - B1.5) the prominent
ionisation stages of silicon are Si {\sc iii \& iv}. In the case of
AV104 \& AV215, the He {\sc ii} lines at 4200 and 4542 \AA\ were
used as further constraints on the temperature, assuming a normal
helium abundance. Given this additional check on T$_{\rm eff}$ the
error in these measurements is $\pm$ 1.0 kK. In AV362 \& AV22, the
coolest objects in our sample, Si {\sc ii \& iii} are the dominant
silicon ions with Si {\sc iv} absent from the spectra. Unfortunately
neither the Si {\sc ii} nor Si {\sc iv} features are very strong in
B2-type stars at low metallicities such as that of the SMC, in fact
we only observe noise features in the region of these lines in AV18.
Although we could not accurately fix an ionisation balance, the
absence of Si {\sc ii \& iv} features provided lower and upper limits,
respectively, for the temperature of AV18. Thus, the error we adopt
in our temperature estimate for AV18 is slightly larger than for the
rest of our sample ($\pm$ 2.0 kK).

The inclusion of sphericity and line-blanketing in the analysis of O-type
dwarfs and supergiants has introduced a new temperature scale for O-type stars
(Martins, Schaerer, \& Hillier 2002; Herrero, Puls, \& Najarro 2002; Crowther
et al. 2002; Repolust, Puls \& Herrero 2003). Compared to plane-parallel,
unblanketed models the temperatures derived for O-type stars are up to 8.0 kK
cooler, when line-blanketed, spherical models are applied. Herrero, Puls, \&
Najarro (2002) noted that this discrepancy decreased to 2.0 kK at the O9.5 Ia
spectral type. For the B0 - B2 -type stars in our sample no notable difference
was observed in our temperature estimates when compared to the calibration of
Dufton et al. (2000) for B-type supergiants (this was also seen in Repolust,
Puls \& Herrero 2003). However for the B3 and B5 star, our temperature
estimates were much lower than their temperature scale. This is probably a
result of our improved Si {\sc ii} model, which for a given temperature,
weakened the Si {\sc ii} line strength. Additionally these two cooler stars
have morphologically different spectra from which their spectral types have
been derived, however they have very similar temperatures. The fact that the
temperatures of these stars are indistinguishable between the two spectral
types, is partly due to the lower luminosity and hence higher gravity of AV22;
this temperature 'degeneracy' is also observed between AV215 and AV104 (B0 \&
B0.5 stars). An additional factor that may be coming into play here is that
AV215 and AV362 both have very high mass-loss rates. Herrero, Puls, \& Najarro
(2002) and Evans et al. (2003) observed that for extreme supergiants the
effective temperature derived is much lower than for normal supergiants. Voels
et al. (1989) also observed a reduction of 2.0 kK in the temperature of
$\alpha$ Cam, an O9.5 Ia type star, with the inclusion of the stellar wind in
the analysis. They attributed this to `wind-blanketing', the backscattering of
radiation onto the photosphere from the stellar wind (see also Repolust, Puls
\& Herrero 2003).

\subsubsection{\it Surface gravity, $\log$ g} 

The logarithmic surface gravity, $\log$ g, is determined in an
iterative process with T$_{\rm eff}$. The sensitivity of the Balmer
line wings to gravity provide an excellent tool for deriving $\log$
g. As H$_{\alpha}$ and H$_{\beta}$ are formed higher up in the
atmosphere than H$_{\gamma}$ and H$_{\delta}$, these lines are more
affected by the wind. Therefore the latter lines were used to
determine the surface gravity with the fitting uncertainty being
typically $\pm$ 0.05 dex. However the uncertainty in the effective
temperature estimates induces an additional though correlated
uncertainty in $\log$ g. These error estimates are presented in
Table~\ref{smcpar1} along with the logarithmic surface gravities,
while the theoretical and observed H$_{\delta}$ profiles are shown
in Fig.~\ref{blue1}. These measured $\log$ g values are in fact
only the effective gravity, which is reduced due to the centrifugal
acceleration by ($v \sin i $)$^2$/R$_{\star}$. The correction to
the derived surface gravities due to the centrifugal force is less
than 0.03 dex for these stars as a result of the low projected
rotational velocities and large stellar radii.  As discussed in
Sect. 2, the measured $v \sin i $ value is an upper limit to the
projected rotational velocity and so the correction to the surface
gravity for these B-type supergiants is negligible. In addition, we
note that the gravity for AV216 may be overestimated due to the
companion star, which is possibly a main-sequence star (see Sect.
2) and would therefore have broader Balmer wings. 

\subsubsection{\it Microturbulence, $\xi$}

Microturbulence, $\xi$, is introduced to ensure no variation of abundance with
line strength from a particular ionic species.  In early B-type stars there is
a sufficient number of O {\sc ii} and Si {\sc iii} lines covering a range of
line strengths from which to determine microturbulence. However, the estimates
from different metal species can differ by up to 10 kms$^{-1}$ (see Vrancken et
al. 1998, 2000; McErlean, Lennon \& Dufton, 1999; Trundle et al. 2002).  This
effect was again observed in our sample where the microturbulence deduced from
the O {\sc ii} lines, $\xi$$_{\rm O}$, were on average 7 kms$^{-1}$ higher than
those derived from the Si {\sc iii} lines, $\xi$$_{ \rm Si}$. Whilst the oxygen
lines have a significant range of EW's from which to determine the
microturbulence, this EW range is not present in one particular multiplet but
over a number of multiplets. Therefore the scatter in oxygen abundances from
one line to another is not simply due to microturbulence but to uncertainties
in the modelling (e.g. errors in the oscillator strengths or departure
coefficients between different multiplets). This is not the case for the Si
{\sc iii} multiplet which has three lines at close wavelength intervals due to
the small difference in the energy levels involved in these transitions, as
such they are effected by the same radiation field. Therefore any errors in the
oscillator strengths or departure coefficients would be similar for each line.
The microturbulent velocities from both elements are presented in
Table~\ref{smcpar1}, although we adopt the microturbulence  derived from the Si
{\sc iii} triplet at 4560 \AA\ (4553, 4568, 4575 \AA). Due to the low T$_{\rm
eff}$ of AV362 and AV22  no O {\sc ii} lines were observed and so $\xi$$_{\rm
O}$ could not be derived (see Fig.~\ref{blue1}). In addition, for AV22 we had
difficulty constraining the silicon microturbulence as the Si {\sc iii} 4575
\AA\ line was not visible in the spectrum and thus there was not a significant
enough range in line strength to determine $\xi$$_{\rm Si}$. Therefore, we
adopted a microturbulence of 10 kms$^{-1}$. A discussion of the effect of
microturbulence  on the abundances of our supergiant sample is given in Sect.
5. 

In previous non-LTE analyses of supergiants the microturbulences
determined have been significantly lower than values derived from
LTE analyses, which tended to be above the speed of sound (Becker
\& Butler 1989; Gies \& Lambert 1992; McErlean et al. 1999;
Vrancken et al. 2000; Trundle et al. 2002). Although for
main-sequence B-type stars the value of $\xi$ is typically lower than
5 kms$^{-1}$, in supergiants the microturbulence is still quite
high even with the inclusion of non-LTE effects. It was suggested
by Lennon et al. (1991) and investigated by Kudritzki et al.
(1992) that the microturbulence may be reduced for supergiants
by the consideration of the wind outflow in these stars. However,
it is evident from the values in Table~\ref{smcpar1} that
microturbulence is still needed in analyses which include
the non-LTE effects and winds of supergiants. In fact the
inclusion of spherical symmetry and treatment of the stellar wind
in our model atmospheres doesn't appear to have any effect on the
magnitude of microturbulence. From a plane-parallel non-LTE
analysis of a group of galactic B-type supergiants, McErlean, Lennon,
\& Dufton (1999) found  an average microturbulence of 11
kms$^{-1}$ and this is consistent with our values. 
For AV216, the giant in our sample, the Si {\sc iii} lines were
compatible with $\xi$ = 0 kms$^{-1}$, while the O {\sc ii} lines implied a $\xi$
= 15 kms$^{-1}$. We note that adopting the latter value would imply a
higher effective temperature estimate for the Si {\sc iii/iv}
ionisation equilibrium. In turn this would lead to He {\sc ii} line
strengths incompatible with the observed helium spectrum of AV216.

\begin{figure*}
\begin{center}
\hspace{-2mm}
\epsfig{file=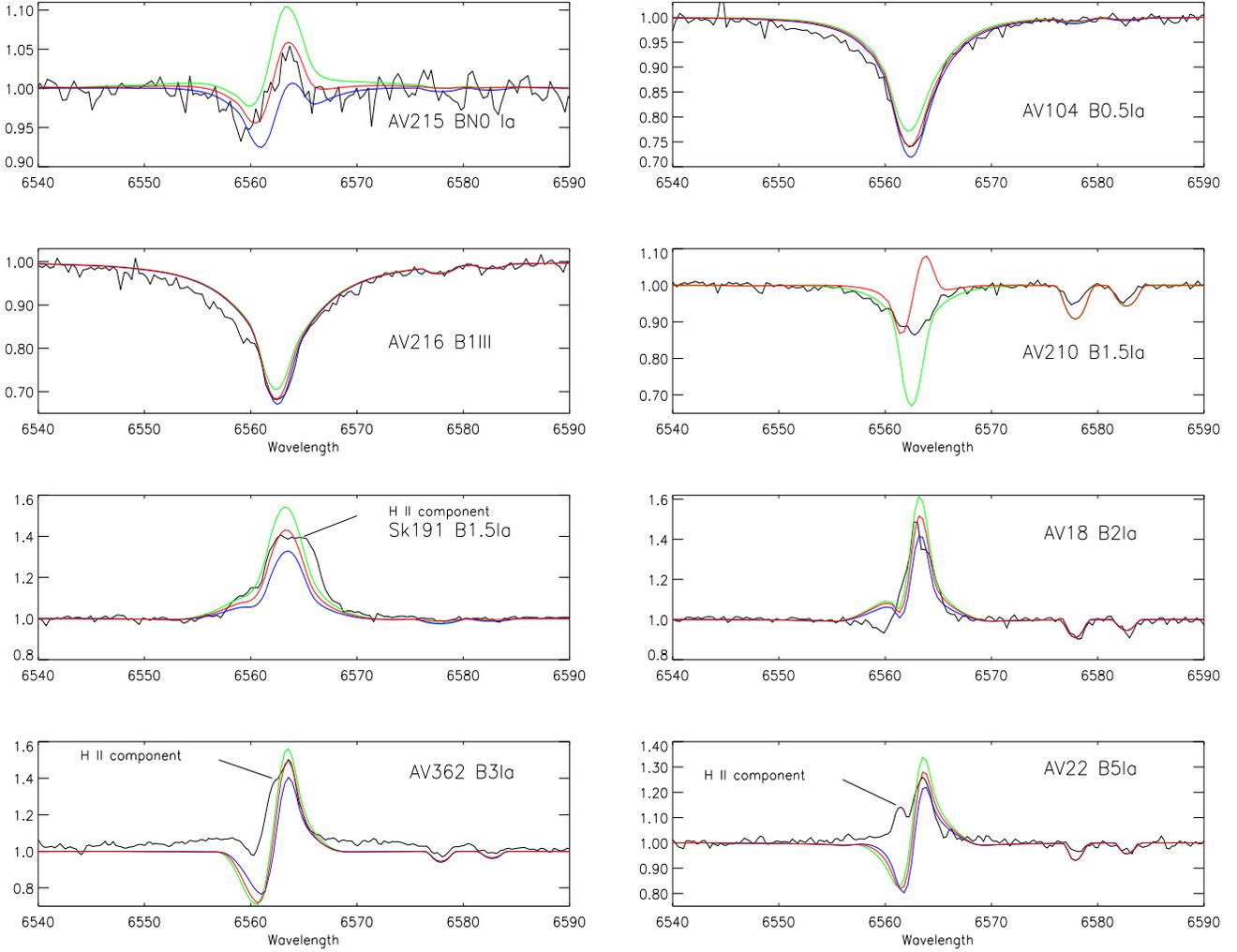, height=140mm, width=180mm, angle=0}
\caption[]
{
H$_{\alpha}$ profile fits for the target stars. Shown are the best fits (red
line) using the parameters presented in Table~\ref{smcpar1} \& \ref{smcpar2}.
The green and blue lines represent a change in mass-loss rate of $\pm$ 15 \%.
However, for AV210, the best fit adopting a $\beta$-parameter of 3 (red line)
and 1 (green line) are  plotted. Note the asymmetry in the H$_{\alpha}$ profile
of AV216, this suggests that the companion star is contaminating AV216's
spectrum. For AV18, Sk191, AV362 \& AV22 the best fit was chosen by fitting the
emission peak. The latter three have nebular contributions in the H$_{\alpha}$
profile which were avoided in the fitting procedure.  Note how for AV362 \& AV22
the models have a large absorption trough which is not present in the observed
spectrum.
}
\label{ha}
\end{center}
\end{figure*}

\subsubsection{\it Radius, luminosity, and mass}

Apparent V-magnitudes were taken, where possible, from the UBVR CCD survey of
Massey (2002). For the brighter targets which were not covered in this survey,
m$_{v}$ was taken from Garmany, Conti, \& Massey (1987). Absolute magnitudes
were than deduced using (B-V)$_{0}$ from the calibration of Fitzpatrick \&
Garmany (1990) and adopting an SMC distance modulus (DM) of 18.9 (Harries,
Hilditch, \& Howarth 2003). As the intrinsic colours of Fitzpatrick \& Garmany
(1990)  were derived for the LMC they are probably not the most suitable values
to apply to the SMC. Yet given the lack of any calibration for the SMC and the
errors involved in these calibrations we find they are the only appropriate
choice. 

The stellar radius was determined using a method described by Kudritzki (1980)

\begin{equation}
5\log R_{\star} = 29.58 - (M_{v} - V) 
\end{equation}

where,

\begin{equation}
V = -2.5 \log  \int_{0}^{\infty} F_{\lambda} S_{v}^{0}(\lambda) d\lambda
\end{equation}
 
F$_{\lambda}$ is the stellar flux from the final theoretical model
of each star (units: erg s$^{-1}$ cm$^{-2}$ \AA$^{-1}$) and
S$_{v}^{0}$($\lambda$) is the V-filter response function for air
mass zero from Matthews \& Sandage (1963). The stellar luminosities
and masses were then calculated using the effective temperatures,
gravities, and radii determined from our analysis.

The error on the stellar radius is dominated by the uncertainty
in our distance estimate. The distance to the SMC has been
determined from many different astrophysical objects (viz.
Cepheids, eclipsing binaries \& RC stars) and has been reviewed
thoroughly in Harries, Hilditch, \& Howarth (2003). From the
various estimates the SMC lies at a DM between 18.7-19.1,
therefore we adopt an error of $\pm$ 0.2. With an additional
error of 0.1 from the adopted apparent magnitudes, this infers an
error of $\pm$ 0.22 on the absolute magnitude estimates giving a
10 \% uncertainty in the stellar radii. This uncertainty in the
radius introduces larger errors in the luminosity and mass
estimates, on the order of 15 \% and 30 \%, respectively. A
systematic error is also introduced to the derived mass-loss
rates due to the uncertainty in the adopted distance to the SMC,
this is discussed further in Sect. 4.2.2.


\subsection{Wind parameters} 

\begin{table*}
\caption[]{
Derived wind parameters for SMC B-type supergiants. Escape
velocities are taken from EVANS04 for AV215, Sk191, AV210 and
AV18. For the other four they are calculated in this work using
the procedure outlined in EVANS04. Terminal velocities,
(v$_{\infty}$), for AV215, Sk191, AV210, and AV18 are taken from
EVANS04, as derived from the SEI method. For the rest of
the sample,  v$_{\infty}$ is calculated by adopting a
v$_{\infty}$/v$_{\rm esc}$  ratio of 2.65 for stars with
T$_{\rm eff}$ $\geq$ 21.0 kK and 1.40 for T$_{\rm eff}$ $<$ 21.0
kK.  The errors quoted are the typical random errors as discussed
in Sect. 4.
}  
\begin{flushleft}
\centering
\begin{tabular}{llccccc} \hline \hline
Star & Spectral & \.{M}& v$_{\infty}$  & v$_{\rm esc}$ & $\beta$ & $\log$(D$_{\rm MOM}$) \\
     & Type     & (10$^{-6}$ M$_{\odot}$ yr$^{-1})$ &(kms$^{-1}$)     &      (kms$^{-1}$) &   & (cgs)    \\        
\hline   
\\            
AV215 & BN0 {\sc I}a  & 1.35 $\pm$ 0.14 & 1400 $\pm$ 100  $^1$ & 447 $^1$ & 1.4 $\pm$ 0.2 & 28.81 $\pm$ 0.10     
\\
AV104 & B0.5 {\sc I}a & 0.40 $\pm$ 0.05 & 1340 $\pm$ 270  $^2$ & 506 $^2$ & 1.0 $\pm$ 0.3 & 28.18 $\pm$ 0.15
\\
AV216 & B1 {\sc III}  & 0.12 $\pm$ 0.02 & 2410 $\pm$ 480  $^2$ & 908 $^2$ & 1.4 $\pm$ 0.3 & 27.86 $\pm$ 0.15
\\
Sk191 & B1.5 {\sc I}a & 0.68 $\pm$ 0.07 &  425 $\pm$ 50 ~~$^1$ & 374 $^1$ & 2.0 $\pm$ 0.2 & 28.11 $\pm$ 0.10
\\
AV210 & B1.5 {\sc I}a & 0.20 $\pm$ 0.05 &  750 $\pm$ 100 ~$^1$ & 290 $^1$ & 3.0 $\pm$ 0.3 & 27.52 $\pm$ 0.15
\\   
AV18  & B2 {\sc I}a   & 0.23 $\pm$ 0.03 &  325 $\pm$ 50 ~~$^1$ & 284 $^1$ & 3.0 $\pm$ 0.2 & 27.66 $\pm$ 0.10
\\
AV362 & B3 {\sc I}a   & 0.80 $\pm$ 0.08 &  270 $\pm$ 50 ~~$^2$ & 192 $^2$ & 1.0 $\pm$ 0.3 & 28.12 $\pm$ 0.15
\\ 
AV22  & B5 {\sc I}a   & 0.23 $\pm$ 0.02 &  280 $\pm$ 60 ~~$^2$ & 198 $^2$ & 1.0 $\pm$ 0.3 & 27.47 $\pm$ 0.15
\\
\hline             
\multicolumn{7}{c}{ Terminal and escape velocities as measured in
$^{1}$ EVANS04 and estimated in $^{2}$ this work.}
\\
\end{tabular}
\end{flushleft}
\label{smcpar2} 
\end{table*} 

In addition to the stellar parameters which describe their
photospheric conditions, the spectra provide
information on their stellar winds. The stellar wind is
parameterised by the mass-loss rate (\.{M}), the exponent of the
velocity law ($\beta$-parameter), and the terminal velocity
(v$_{\infty}$). The former two parameters are derived here
from the H$_{\alpha}$ profile and the latter from UV P-Cygni
profiles.

\subsubsection{Terminal velocity, v$_{\infty}$}

Terminal velocities of AV215, Sk191, AV210 \& AV18 have been presented in
EVANS04, where a more detailed discussion can be found. Briefly, v$_{\infty}$
was derived from the UV P Cygni lines using the method described by Haser et
al.  (1995).  This procedure utilises the Sobolev Exact Integration (SEI)
method introduced by Lamers, Cerruti-Sola, \& Perinotto (1987; see also Hamann
1981). For early B-type stars the doublets Si {\sc iv} 1394, 1403 and C {\sc
iv} 1548, 1551 display the most prominent wind features and thus were the
diagnostic lines used by EVANS04, see the fits to  these profiles in Fig. 2 of
EVANS04.

Unfortunately four stars in our sample showed no significant wind
features in the aforementioned UV lines; AV104, AV216, AV362 \&
AV22 (EVANS04). Therefore terminal velocities for these stars
were estimated by an alternative method. Radiation driven wind
theory (Castor, Abbott, \& Klein 1975; Abbott 1978) predicts that
terminal and escape velocities scale together as a function of
$\alpha$, such that $v_{\infty}/v_{\rm esc} \sim {\hat{\alpha}}/(1 -
\hat{\alpha})$, if the finite cone-angle subtended by the stellar
disk is accounted for. $\hat{\alpha}$ is dependent on the
effective temperature and is defined as the effective value of the
force multiplier parameter, $\alpha$, defining the exponent of
the line strength distribution function. Lamers, Snow, \&
Lindholm (1995) found, for stars with temperatures $\geq$ 21 kK,
that the v$_{\infty}$/v$_{\rm esc}$ ratio was approximately 2.6
but that at lower temperatures ($<$ 20 kK) this dropped to 1.3.
Kudritzki \& Puls (2000) found, from a comparison of the Lamers
sample of 68 stars with an additional set of stars ($\sim$ 200)
from Prinja \& Massa (1998), that above 21 kK this ratio is 2.65
and below 21 kK it is 1.4. 

In EVANS04 a comparison of v$_{\infty}$/v$_{\rm esc}$ derived from the Lamers
and Prinja \& Massa samples discussed above with those derived from SMC OB-type
supergiants, showed no convincing variation in this relationship due to the
different metallicity environments. The average v$_{\infty}$/v$_{\rm esc}$ from
the SMC OB-type supergiants with T$_{\rm eff}$ $>$ 21 kK is within the scatter
of the Galactic star results. Therefore in order to derive the terminal
velocities of AV104, AV216, AV362 \& AV22 we have adopted the
v$_{\infty}$/v$_{\rm esc}$ ratio of 2.65 for the former two stars and 1.4 for
the two latter, cooler stars. The escape velocities were derived using the same
procedure as described in EVANS04 (see equations 1 \& 2), where v$_{\rm esc}$
was calculated from the ratio of stellar mass to radius and corrected for the
radiative acceleration due to electron scattering. We should note that in all
three samples (Lamers, Snow, \& Lindholm 1995; Prinja \& Massa 1998; EVANS04)
there is a large scatter of approximately 20 \% in the v$_{\infty}$/v$_{\rm
esc}$ at a given temperature. This is reflected in the the error estimates for
v$_{\infty}$ given in Table~\ref{smcpar2} for the four stars where we have used
this method.

\subsubsection{Mass-loss rate \& $\beta$-parameter}

\begin{table*}
\caption[]{
Derived non-LTE absolute abundances for SMC B-type supergiants.
Abundances are given as $\log$ [N(X)/N(H)] + 12, where X
represents the appropriate element. Numbers in the parenthesis
represent the number of lines used in the abundance analysis. The
errors represent the standard deviation of the mean and account
for systematic errors as discussed in Sect. 5.
}  
\begin{flushleft}
\centering
\begin{tabular}{lcccccccc} \hline \hline
Star &  C {\sc ii}~~~~~~~~~~~~~n & N {\sc ii}~~~~~~~~~~~~~~n & O {\sc ii}~~~~~~~~~~~~~~n & Mg {\sc ii} ~~~~~~~~~~n & Si {\sc ii}~~~~~~~~~~~~n & Si {\sc iii}~~~~~~~~~~~n & Si {\sc iv} ~~~~~~~~~~~n 
\\
\hline 
\\     
AV215 & 6.91 $\pm$ 0.10  (1) & 7.96 $\pm$ 0.12 (3) & 7.97 $\pm$ 0.14 ~(6) &$<$7.20~~~~~~~~~    (1) & 		     & 7.10 $\pm$ 0.15 (3) & 7.14 $\pm$ 0.12 (1)
\\
AV104 & $<$7.00~~~~~~~~~  (1) & 7.41 $\pm$ 0.15 (2) & 8.05 $\pm$ 0.32 (12) & 6.90 $\pm$ 0.05 (1) &	             & 6.70 $\pm$ 0.11 (3) & 6.67 $\pm$ 0.22 (1) 
\\
AV216 & 6.76 $\pm$ 0.16  (1) & 7.14 $\pm$ 0.20 (1) & 8.22 $\pm$ 0.39 ~(7)  &6.64 $\pm$ 0.08 (1) &		     & 6.66 $\pm$ 0.22 (3) & $<$6.71 ~~~~~~~~~(1)
\\
Sk191 & 6.89 $\pm$ 0.10  (1) & 7.63 $\pm$ 0.19 (6) & 8.20 $\pm$ 0.18 (12) & 6.98 $\pm$ 0.06 (1) &		     & 6.75 $\pm$ 0.16 (3) & 6.50 $\pm$ 0.25 (1)
\\
AV210 & 6.93 $\pm$ 0.14  (3) & 7.60 $\pm$ 0.12 (7) & 8.18 $\pm$ 0.18 (12) & 6.89 $\pm$ 0.07 (1) &		     & 6.69 $\pm$ 0.14 (3) & 6.70 $\pm$ 0.11 (1)
\\
AV18  & 6.98 $\pm$ 0.18  (3) & 7.50 $\pm$ 0.30 (7) & 8.26 $\pm$ 0.21 (12) & 6.75 $\pm$ 0.10 (1) & $<$7.03 ~~~~~~~~   (2) & 6.81 $\pm$ 0.11 (3) &
\\
AV362 & 7.12 $\pm$ 0.48  (3) & 8.22 $\pm$ 0.25 (3) &		  & 6.72 $\pm$ 0.06 (1) & 6.60 $\pm$ 0.13 (2) & 6.66 $\pm$ 0.19 (3) &
\\
AV22  & 7.07 $\pm$ 0.27  (3) & 7.92 $\pm$ 0.16 (3) &		  & 6.83 $\pm$ 0.09 (1) & 6.75 $\pm$ 0.16 (2) & 6.61 $\pm$ 0.15 (2) &
\\ 
\hline 
\end{tabular}
\end{flushleft}
\label{smcabund}
\end{table*}

The mass-loss rate is derived mainly from the morphology of the H$_{\alpha}$
profile, with less weight given to H$_{\beta}$ profile. This procedure has been
described in detail in Puls et al. (1996) and Kudritzki et al. (1999). The
$\beta$-parameter effects the central emission core of the H$_{\alpha}$ profile
and also its FWHM. Therefore for thick winds (i.e when there is emission in the
H$_{\alpha}$ profile), the $\beta$-parameter and mass-loss rate can be obtained
simultaneously.  The derived mass-loss rates are presented in
Table~\ref{smcpar2}, the uncertainties quoted are the random errors involved in
the fitting process which are typically in the range 10 - 15 \%. In addition we
should consider the systematic error incurred through the adopted distance
modulus. The uncertainty in DM follows through to the mass-loss rate from the
10 \% uncertainty it induces on the stellar radius. If we assume that we get
identical profiles as long as Q is constant (i.e. that we are fitting Q and not
\.{M}), where Q = \.{M}/R$_{\star}^{3/2}$, then the systematic error in the
derived mass-loss rates is $\pm$ 15 \% (Puls et al. 1996). Fits to the
H$_{\alpha}$ profiles are presented in Fig.~\ref{ha}, where we show the best
fits (red line) along with models representing a change in \.{M} of $\pm$ 15 \%
(green and blue lines). Some nebular contamination was detected in Sk191, AV362
\& AV22 from comparisons between the {\sc uves} and {\sc emmi} datasets. After
careful extraction of the nebular contribution to the H$_{\alpha}$ profiles the
{\sc emmi} data were considered more reliable for AV362 \& AV22 whilst for
Sk191 the {\sc uves} data appeared less contaminated. In fitting the
H$_{\alpha}$ line of these stars any obvious nebular component was avoided (see
Fig.~\ref{ha}). 

For those stars with thin winds, (AV104 \& AV216), H$_{\beta}$ was used as a
constraint on the $\beta$-parameter, whilst the core depth of the H$_{\alpha}$
line constrained \.{M} (see Fig. ~\ref{ha}). For too high a $\beta$-parameter,
an emission peak appears in the H$_{\beta}$ theoretical profile which is not
observed. This method provides an upper limit for the $\beta$-parameter. 

In the case of AV215 (BNO Ia), the profile shape of the H$_\alpha$ line
provided a means of determining both \.{M} and $\beta$ simultaneously. However,
the models could not simultaneously reproduce the depth and width of the
absorption core. Kudritzki et al. (1999) noticed this problem in a B1 Iab-type
supergiant, HD13854, suggesting it may be a lack of line-blanketing which was
affecting the stratification of the source function of H$_\alpha$ throughout
the wind. This is not supported by our analysis as it has been carried out with
a version of {\sc fastwind} that includes an approximation of the
line-blanketing and blocking. This inability to reproduce the depth of the
blueward absorption core is also seen for AV18 (B2 Ia). This star is similar to
the B2 Ia-type stars HD 41117 \& HD 14143 discussed by Kudritzki et al. (1999),
in that its emission peak could be fitted giving \.{M} and $\beta$, whilst the
absorption core could not be matched.

The H$_\alpha$ profiles of the two B1.5 Ia type stars illustrate the extreme
differences between the winds of stars with the same spectral and luminosity
classification; whilst Sk191 is in emission, AV210 is in absorption. In
Fig.~\ref{ha} we show two fits to AV210, one with $\beta$ = 1 and the other
with $\beta$ = 3. Although, these are extremely different profiles the
mass-loss rate  only varies by 25 \%. We have adopted the parameters from the
$\beta$ = 3 case as the observed H$_{\beta}$ line is reproduced and the
H$_{\alpha}$ absorption core depth,  though not the profile shape, can also be
matched.

In the coolest stars, AV362 \& AV22, \.{M} and $\beta$ were derived from the
height and width of the H$_\alpha$ emission. Note in Fig.~\ref{ha} that the 
code predicts large absorption troughs not observed in the actual data, yet
they are an inevitable consequence of the Lyman continuum and the Lyman
resonance line becoming optically thick at lower temperatures and large
mass-loss rates (e.g. Kudritzki \& Puls 2000). Although there is probably some
nebular contamination in the H$_\alpha$ profile of both stars, it is unlikely
to have masked such a large absorption core. One can speculate that this
problem might be resolved by including the effects of clumping in the analysis,
since the mass-loss rates derived here would be overestimated and the Lyman
optical depths might be reduced. We will come back to this again in Sect. 6.3
however a detailed discussion of the effects of clumping is beyond the scope of
this work.

\section{Chemical Abundance Analysis}           

\begin{table*}
\caption[]{
Comparison of mean abundances for the SMC B-type supergiants with previous
studies of abundances in the SMC. Along with our mean B-type supergiant
abundances as derived from Table~\ref{smcabund} we show the mean abundances of
A-type supergiants from Venn et al. (1999, 2003b; Si corrected for non-LTE
effects.), SMC main-sequence B-type star AV304 (Rolleston et al. 2003;
corrected for non-LTE effects), NGC330 cluster main-sequence B-type stars
(LEN03; corrected for non-LTE effects), and SMC H {\sc ii} regions from Kurt et
al. (1999).  The errors represent the standard deviation of the mean. 
}  
\begin{flushleft}
\centering
\begin{tabular}{lccccc} \hline \hline
        & \multicolumn{2}{c}{Supergiants}  & \multicolumn{2}{c}{B stars}& H {\sc ii} 
\\
Metal & This work & A-type &  NGC330 & AV304 & regions
\\
\hline
\\
C   & 7.30 $\pm$ 0.09 &                 & 7.26 $\pm$ 0.15 & 7.41 $\pm$ 0.18 & 7.53 $\pm$ 0.06
\\
N   & 7.67 $\pm$ 0.27 & 7.52 $\pm$ 0.10 & 7.51 $\pm$ 0.18 & 6.55 $\pm$ 0.01 & 6.59 $\pm$ 0.08
\\
O   & 8.15 $\pm$ 0.07 & 8.14 $\pm$ 0.06 & 7.98 $\pm$ 0.13 & 8.16 $\pm$ 0.33 & 8.05 $\pm$ 0.05
\\
Mg  & 6.78 $\pm$ 0.16 & 6.83 $\pm$ 0.08 & 6.59 $\pm$ 0.14 & 6.73            & 
\\
Si  & 6.74 $\pm$ 0.11 & 6.92 $\pm$ 0.15 & 6.58 $\pm$ 0.32 & 6.74 $\pm$ 0.03 & 6.70 $\pm$ 0.20
\\
\hline 
\\
\end{tabular}
\end{flushleft}
\label{smccomp} 
\end{table*} 

In previous investigations of the atmospheres of supergiants, their
has been clear evidence of chemical processing (McErlean, Lennon, \&
Dufton 1999). The photospheres of these luminous stars have chemical
patterns which indicate contamination of the stellar surface by the
products of CN-cycle burning and thus we have carried out an
abundance analysis on this stellar sample to study this  further. 
Adopting the atmospheric parameters presented in the previous
section we derived the chemical composition of our stellar sample.
The results of the absolute abundance analysis are presented in
Table~\ref{smcabund}. We have also compared our mean abundances to
those derived for B-type giants and main-sequence stars in NGC330,
an unprocessed B-type star AV304 and SMC H {\sc ii} regions, see
Table~\ref{smccomp}. 

The most dominant uncertainty on B-type supergiant abundances is the
adopted microturbulence. As discussed in Sect. 4.1.3 each element
implies a different value for the microturbulence (see
Table~\ref{smcpar1}). As we have adopted the microturbulence from the
Si {\sc iii} lines, $\xi_{\rm Si}$, it is necessary to explore the
effect of this choice of microturbulence on the other elements. Firstly
the uncertainty in $\xi$$_{\rm Si}$ is $\pm$ 2 kms$^{-1}$, which
introduces an error of $\sim$ 0.05 for weak lines and typically $\pm$
0.10 for strong lines. For example the carbon abundance, which depends
mainly on the weak C {\sc ii} feature at 4267 \AA\ only varies by $\pm$
0.02 for the B0 - B1.5 stars. As this line becomes stronger for the B2
- B5 -type stars the effect of microturbulence is larger increasing to
$\pm$ 0.07. The largest effects of the uncertainty in $\xi$$_{\rm Si}$
are on the abundances of Si {\sc iv} in AV215 \& AV104 and nitrogen in
AV362 \& AV22. If we adopt the microturbulence from the oxygen lines,
there is a decrease in the abundance of carbon, nitrogen, and magnesium
of normally less than 0.1 dex. However the silicon and oxygen
abundances are decreased by up to 0.2 dex depending on the degree of
change in the microturbulence and the strength of the spectral
features.

\subsection{Helium} 

Achieving a consistent fit to both the singlet and triplet neutral
helium lines in a stellar spectrum is a known problem. Voels et al.
(1989) concluded that the extension of the stellar atmosphere was
causing this effect and could be resolved by the inclusion of
sphericity in the model atmospheres. Studies of OB-type supergiants
which omit microturbulence tend to find very high helium fractions,
of up to 0.4, in order to get a fit to all He {\sc i} lines (Lennon
et al. 1991; Herrero et al. 1992). By including a non-zero
microturbulence McErlean, Lennon, \& Dufton (1998) found that a more
consistent, although not perfect, fit to the helium lines could be
achieved with a solar or near solar helium fraction of 0.09. Due to
the difficulty in constraining the helium abundance from the neutral
helium lines we have adopted a solar helium fraction of 0.09. No
significant variation from this helium abundance was noted in our
sample, although consistent fits to the helium spectra were not
always achieved. We should note that the helium abundance has an
indirect effect on the temperature and gravity and hence the
mass-loss rate. McErlean, Lennon, \& Dufton (1998) showed that by
changing the helium fraction from 0.1 to 0.2, the T$_{\rm eff}$ and
$\log$ g were decreased by 0.5 kK and 0.05 dex, respectively, which
would change \.{M} by less than 10 \%.

\subsection{Carbon, nitrogen \& oxygen}

The carbon abundances are based primarily on the C {\sc ii} 4267 \AA\ line,
with the 6580 \AA\ multiplet being visible for the B2 - B5 stars (see
Fig.~\ref{ha}).  From Table~\ref{smcabund} we see that the mean carbon
abundance, from those stars for which it is derived solely from the 4267 \AA\
line (B0 - B1.5), is $\pm$ 0.1 dex lower than that determined by including the
6580 \AA\ multiplet (B2 - B5). In the coolest stars the 6580 \AA\ multiplet
actually gives abundances of $\sim$ 0.25 dex higher than those determined from
the 4267 \AA\ line. The 4267 \AA\ line is known to be particularly sensitive to
non-LTE effects (Eber \& Butler 1988), resulting in lower abundances than
derived from other carbon lines (viz. 3919, 3921, 4650 \AA). This has been
found by LEN03 in the non-LTE analyses of B-type stars in NGC330 and that of an
unprocessed B-type star AV304 (LEN03; Rolleston et al. 2003). However the
carbon abundances for the stars in our sample show little scatter and have a
mean abundance for the sample of 6.96 dex.

The mean oxygen abundance is in agreement with these previous studies, the
large errors in Table~\ref{smcabund} represent the uncertainty caused by
adopting $\xi$$_{\rm Si}$. No oxygen abundance could be determined for the
AV362 \& AV22, as these are too cool for the presence of visible oxygen lines
in their spectra.  

The nitrogen abundances for our sample exhibit a large scatter with the mean
being 7.67 dex, a factor of 13 more overabundant than AV304.  

\subsection{$\alpha$-process elements; magnesium \& silicon}

Magnesium abundances are based solely on the Mg {\sc ii} 4481 \AA\
doublet but are reasonably  consistent throughout our sample. For
AV215, the magnesium line was weak and blended with a He {\sc i}
line, thus providing us with only an upper limit.

The silicon abundance is very sensitive to the adopted
temperature and this is clear from the range of abundances
presented in Table~\ref{smcabund}. However, the mean abundance
agrees well with those of AV304 and SMC H {\sc ii} regions.

\section{Discussion}

\subsection{Chemical composition of SMC}

Before analysing our samples composition, it is useful to discuss the baseline
composition of the SMC to which we can relate our absolute abundances.
Rolleston et al. (2003) showed that the main-sequence star AV304 has a chemical
composition similar to that derived from SMC H {\sc ii} regions. Since this
star shows no signs of processing in its photosphere it should represent the
present-day composition of the SMC. However this was an LTE analysis and in
order to compare our non-LTE abundances with those of AV304 we need to take
into account the non-LTE effects. LEN03 carried out such corrections on AV304
and their NGC330 B-type star sample. In addition they applied corrections to
the carbon abundance to account for the lower abundances derived from the 4267
\AA\ line. As mentioned in Sect. 5.2 the absolute carbon abundance is effected
by the problematic line 4267 \AA, which gives abundances lower than those
derived by weaker C {\sc ii} lines. Thus a direct comparison of the carbon
abundance derived from 4267 \AA\ and that from H {\sc ii} regions is
inappropriate.  Gies \& Lambert (1992) and Vrancken et al. (2000), derived
carbon abundances from weak lines and the 4267 \AA\ line, respectively. From a
comparison of these studies LEN03 found the weak lines gave abundances which
were 0.34 dex higher than those determined from the 4267 \AA\ line and
subsequently applied this correction to the carbon abundances of AV304 and the
NGC330 stars. We have also applied this correction to our derived carbon
abundances which rely mainly on the 4267 \AA\ line. In Table~\ref{smccomp} we
present the corrected AV304 abundances that can be used as a baseline SMC
chemical composition along with the SMC H {\sc ii} region abundances (Rolleston
et al. 2003; Kurt et al. 1999). Note the excellent agreement between  AV304 and
the H {\sc ii} regions. 

Also included in Table~\ref{smccomp} are the mean abundances from our B-type
supergiant sample, from SMC A-type supergiants(Venn et al. 1999, 2003b) and the
corrected abundances of NGC330 B-type stars (LEN03). The Si abundance presented
in Venn et al. (1999) was derived from LTE calculations, corrections similar to
those carried out for AV304 and the NGC330 stars have been applied to this
value to account for non-LTE effects. Comparing the abundances of the
$\alpha$-process elements (Mg \& Si) in each set of objects we see, as
expected, that they are in reasonable agreement indicating no processing of
these elements. In fact assuming that the galactic standards of LEN03 and
Rolleston et al. (2003) represent the galactic B-type star abundances, we find
that the $\alpha$-processed elements in our sample are 0.42 dex lower than the
galactic objects. This is in agreement with the 0.44 dex difference found
between HR2387 and AV304 (Rolleston et al. 2003).

The most significant result is the variation in the nitrogen abundances. Whilst
the AV304 nitrogen abundance and that of the H {\sc ii} regions agree, those
for the B \& A-type supergiants and the NGC330 stars are a factor of 9 to 14
greater. It is interesting to note that within the B-type supergiant sample
itself, a sizeable variation in nitrogen abundance is also present (see
Table~\ref{smcabund}). These enhancements are much greater than any error
incurred in our analysis and are therefore considered to be real. The most
likely explanation for such an enhancement of nitrogen is that we are seeing
the CNO processed material contaminating the chemical composition of the
photosphere. If this is the case, with such high nitrogen abundances we might
expect a corresponding but smaller enhancement in helium with depletions of
carbon and possibly oxygen, keeping the sum of CNO nuclei constant.  However a
comparison of the oxygen abundance with that of AV304 shows excellent
agreement, with no indication of an underabundance present. On the other hand
carbon appears to be underabundant by 0.11 and 0.23 in comparison to AV304 and
the H {\sc ii} regions, respectively. This is comparable to the carbon
depletions produced in the Maeder \& Meynet (2001) stellar evolution models
from similar nitrogen enhancements ($<$ 0.25 dex). Nevertheless given the
uncertainties in the carbon abundance and the magnitude of the correction
applied to carbon due to the 4267 \AA\ line, this slight underabundance may not
be significant. The CNO pattern of our B-type supergiant sample is in good
agreement with that of the A-type supergiants and NGC330 B-type stars i.e. N is
enhanced, whilst C \& O appear to be of normal SMC metallicity.

\subsection{Evolutionary status}

\begin{figure}
\begin{center}
\epsfig{file=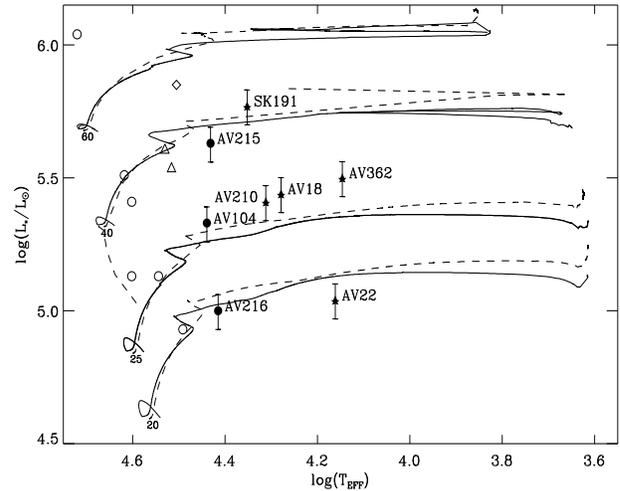, height=70mm, width=88mm, angle=0}
\caption[]
{
HR-diagram, luminosity as a function of temperature. Shown are the stellar
evolution tracks of Maeder \& Meynet (2001) for an assumed initial rotational
velocity of 0 (---; solid lines) and  300 kms$^{-1}$ (- -; dashed lines) at
stellar masses of 20, 25, 40, \& 60 M$_{\odot}$. Included are the results from
our early and mid B-type supergiants ($\bullet$; $\star$),with errors in
luminosity representing $\pm$ 15 \% , SMC O-type supergiants of Hillier et al.
($\triangle$ ; 2003) and Crowther et al. ($\diamond$ ; 2002), and the  SMC
O-type dwarfs of Bouret et al. ($\circ$ ; 2003)
}
\label{hrdiag}
\end{center}
\end{figure}

\begin{figure}
\begin{center}
\epsfig{file=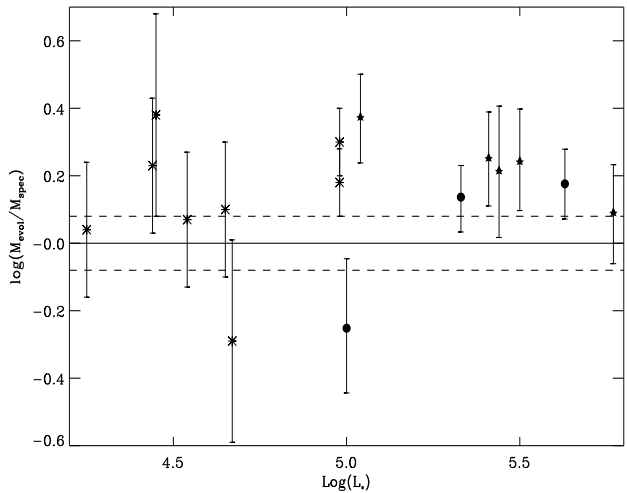, height=70mm, width=88mm, angle=0}
\caption[]
{
Comparison of evolutionary and spectroscopic masses as a function of
luminosity. Included are the results from our early and mid B-type supergiants
($\bullet$; $\star$) and NGC330 B-type stars from Lennon et al. ($\ast$ ; 2003).
The error bars represent the error in the surface gravity. Also shown is the
uncertainty in mass due to the adopted distance modulus (- -).
}
\label{mass}
\end{center}
\end{figure}

\begin{figure}
\begin{center}
\epsfig{file=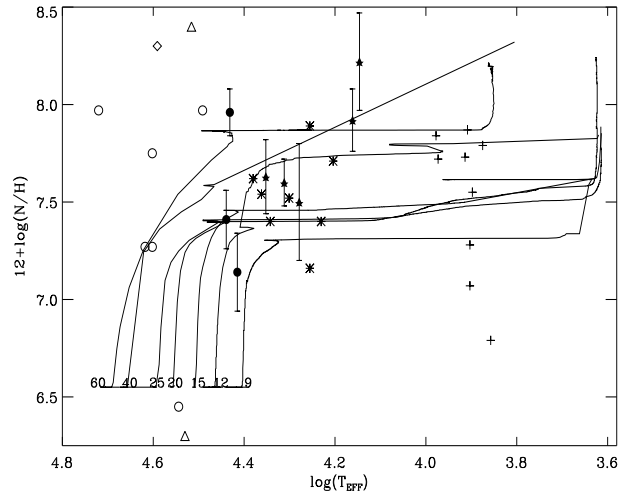, height=70mm, width=88mm, angle=0}
\caption[]
{
Surface nitrogen abundances as a function of temperature. Shown are the stellar
evolution tracks of Maeder \& Meynet (2001) for an assumed initial rotational
velocity of 300 kms$^{-1}$ and various stellar masses. Included are the results
from our early and mid B-type supergiants ($\bullet$; $\star$), SMC O-type
supergiants of Hillier et al. ($\triangle$ ; 2003) and Crowther et al.
($\diamond$ ; 2002), SMC O-type dwarfs of Bouret et al. ($\circ$ ; 2003),
NGC330 B-type stars of Lennon et al. ($\ast$ ; 2003), and A-type supergiants from
Venn et al. (+; 2003b). The error bars illustrate the random and systematic
errors on the nitrogen abundance. 
}
\label{N_teff}
\end{center}
\end{figure}

\begin{figure}
\begin{center}
\epsfig{file=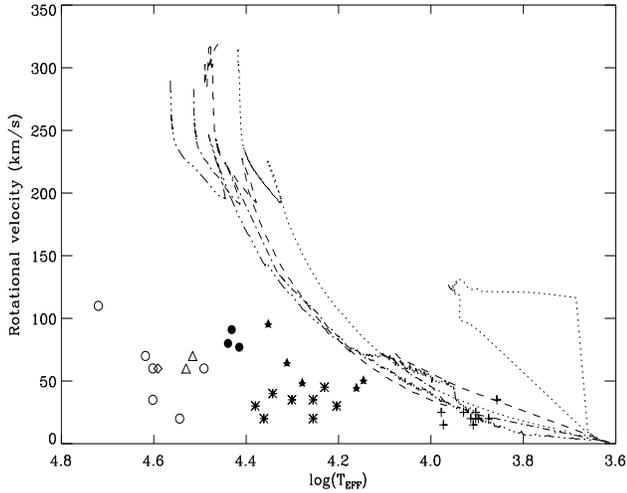, height=70mm, width=88mm, angle=0}
\caption[]
{
Rotational velocities as a function of temperature. Shown are the stellar
evolution tracks of Maeder \& Meynet (2001) for an assumed initial rotational
velocity of 300 kms$^{-1}$ and various stellar masses. Included are the $v \sin i$ 
values from our early and mid B-type supergiants ($\bullet$; $\star$),
SMC O-type supergiants of Hillier et al. ($\triangle$ ; 2003) and Crowther et
al. ($\diamond$ ; 2002), SMC O-type dwarfs of Bouret et al. ($\circ$ ; 2003),
NGC330 B-type stars Lennon et al. ($\ast$ ; 2003), and A-type supergiants from
Venn et al. (+; 1999). 
} 
\label{vrot_teff}
\end{center}
\end{figure}

To help understand the evolutionary status of these B-type
supergiants/giant we have compared our results to the stellar
evolution tracks of Maeder \& Meynet (2001) which include
rotationally induced mixing (see Figs.~\ref{hrdiag}, \ref{N_teff},
\& \ref{vrot_teff}). As noted by LEN03 the initial abundances of
these models are scaled to solar abundances and so by adopting 0.2
Z$_{\odot}$ they greatly overestimate the initial nitrogen abundance
i.e. instead of an SMC nitrogen abundance of 6.55, they use
one-fifth solar or 7.3 dex. We have recalibrated these models to SMC
nitrogen abundances, following the discussion of LEN03, who assumed 
the initial nitrogen abundance has no effect on the amount of excess
nitrogen produced and simply rescaled it to the SMC metallicity.

The position of our stellar sample on the HR-diagram is shown in
Fig.~\ref{hrdiag}, overlayed on the rotating (- -) and non-rotating (---)
tracks of  Maeder \& Meynet (2001). Note that all our stars lie redward of the
main-sequence phase, with the two coolest stars having positions consistent
with the core-helium burning phase (this phase is represented by the bump at
$\log$ $T_{\rm eff}$ = 4.2 - 4.3). By interpolating between the rotating
stellar evolution tracks we estimated the evolutionary masses (M$_{evol}$) of
the sample, which  are listed in Tables~\ref{smcpar1}. Figure~\ref{mass}
illustrates the difference between M$_{evol}$ and the derived spectroscopic
masses, M$_{spec}$, which were determined from the stellar radii and surface
gravities (see Table~\ref{smcpar1}). This plot implies a mass discrepancy where
M$_{evol}$ $>$ M$_{spec}$, for most of the sample. LEN03 suggested that this
mass discrepancy, could be accounted for by the uncertainties in the surface
gravity which were considerably larger in their work (due to low S/N). Moreover
the spectroscopic masses are dependent on the distance modulus (DM) which
introduces an additional systematic error. In Fig.~\ref{mass} the solid and
dashed line represents the zero point of $\log$ (M$_{evol}$/M$_{spec}$) and the
errors on this value due to the adopted DM. The errors in the gravity and
distance modulus can account for some of this mass discrepancy but there
remains an inconsistency of 0.1 - 0.2 dex. We note that one of our stars,
AV216, has M$_{evol}$ $<$ M$_{spec}$, similar to one star from LEN03. As
discussed in Sect. 4.1.2 we may have overestimated the surface gravity of this
star due to the presence of a companion and this would, in turn, lead to an
overestimation of M$_{spec}$.

In Fig.~\ref{N_teff} we plot the recalibrated stellar evolution tracks, showing
the development of the surface nitrogen abundances as a function of
temperature. We have included the photospheric nitrogen abundances of all the
SMC OBA-type supergiants and dwarfs available in the literature (Hillier et al.
2003; Crowther et al.  2002; Bouret et al. 2003; LEN03; Venn et al. 2003b) as
well as the results from our B-type supergiants. The initial rotational
velocity of the stellar models which can reproduce nitrogen abundances similar
to those observed needs to be at least 300 kms$^{-1}$ and these models show no
significant slowdown by the end of the hydrogen-burning stage. Although these
rapidly rotating models imply significant nitrogen enhancements some of the
OBA-type supergiants have abundances up to a factor of five larger than
predicted. 

Fig.~\ref{vrot_teff} shows the observed projected rotational velocities of the
OBA-type supergiants and dwarfs against the predicted rotational velocities.
Note that we are probably overestimating the values for the  supergiants due to
spectral line broadening being dominated by macroturbulence (see Sect. 2;
Howarth et al. 1997; Ryans et al. 2002). The observed velocities are
significantly lower than expected from the stellar evolution models. In order
for these stars to have rotational velocities similar to those predicted, the
inclination of the stellar rotation axis would have to be $<$ 20$^{\circ}$.
Assuming that the distribution of the rotational axis, i, is  random, it is
highly unlikely that our sample of stars have such large rotational velocities
and that we are observing them all pole on.  

The O-type dwarfs of Bouret et al. (2003) have masses similar to the masses of
our sample, as such we will consider them to be representative of the
progenitors for the B-type supergiants. Since these O-type stars show a similar
range of nitrogen enhancements to the B-type supergiants, it is reasonable to
assume that the additional nitrogen is produced while the stars are still on
the main-sequence. If the nitrogen enhancement is truly a product of
rotationally induced mixing, this process needs to be more efficient at lower
rotational velocities than predicted by Maeder \& Meynet (2001), see also the
discussion of Herrero \& Lennon (2002).

A factor that needs to be carefully considered in stellar evolution models is
the effect of the mass-loss rate. This is incorporated in the Maeder \& Meynet
(2001) models and is responsible for the loss of angular-momentum through the
stellar wind. These models use mass-loss rates from Kudritzki \& Puls (2000)
and are modified by rotation using an expression dependent on $\alpha$, a force
multiplier parameter. Maeder \& Meynet assume that $\alpha$ = 0.6 but as
discussed in Sect. 6.3 this quantity is not reliably established for B-type
stars. Additionally Kudritzki \& Puls (2000) used only unblanketed analyses of
the O and B type supergiants, whilst more recent analyses will give more
accurate mass-loss rates due to their inclusion of line-blanketing in the model
atmosphere codes (see Sect. 6.3 for a further discussion). Clearly some
allowance for the variation of mass-loss rates as a function of metallicity
should also be included in the stellar evolution models.

Binary star evolution models by Wellstein, Langer, \& Braun (2001), can also
produce nitrogen enriched blue supergiants with similar temperatures and
luminosities to the SMC supergiants in our sample. Through mass transfer from
the primary star, the low mass secondary in a binary system becomes more
massive. These stars may appear to have lower spectroscopic masses than
expected from their luminosity. Thus this could account for some of the mass
discrepancy in our stellar sample. One advantage of the binary evolution
models, is that they can place these blue supergiants on the post main-sequence
gap which presently can not be achieved with single-star models due to their
rapid evolution through this part of the HR-diagram.  Since the primary of the
binary system will evolve into a low mass helium star by the time the secondary
is a blue supergiant, it is unlikely that the binary systems could be detected
through radial velocity variations. Wellstein, Langer, \& Braun (2001) predict
radial velocity variations of $\sim$ 15 kms$^{-1}$.


\begin{figure}
\begin{center}
\epsfig{file=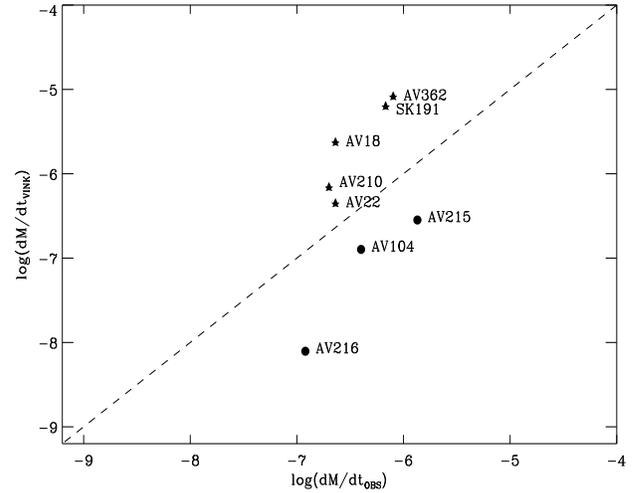, height=70mm,width=88mm}
\caption[]
{
Comparison of theoretical and observed mass-loss rates.
Theoretical predictions are calculated using the metallicity
dependent mass-loss recipes of Vink et al. (2001) for Z =
0.2 Z$_{\odot}$. The solid circles ($\bullet$)
represent the SMC early B-type supergiants, 
whilst the solid stars ($\star$) represent the mid B-type supergiants.
}
\label{vink}
\end{center}
\end{figure}

\subsection{Winds of B-type supergiants}

According to radiative line-driven wind theory, the wind-momentum
and hence mass-loss rate should be dependent on metallicity and
may be expressed as a function of the force multiplier parameter,
$\alpha$. This parameter is the ratio of the optically thick line
acceleration to the total line acceleration and the
metallicity dependence of the mass-loss rate is found to scale as

\begin{equation}
 \indent
 \dot{M} \sim (Z_{*}/Z_{\odot})^{\frac{1 - \alpha}{\alpha^{\prime}}}
\end{equation}

where, $\alpha^{\prime}$ = $\alpha$ - $\delta$ and $\delta$ is a
force multiplier parameter that accounts for the ionisation balance
in the wind (Kudritzki et al. 1989; Puls, Springmann, \& Lennon
2000). Theoretically, this metallicity dependence is predicted to be
in the range $Z^{0.5} - Z^{0.85}$ for OB-type supergiants
(Kudritzki, Pauldrach, \& Puls 1987; Leitherer, Robert, \& Drissen
1992; Vink et al. 2001). Observationally,  a significant enough 
sample of stars has not yet been analysed to get a reliable estimate
of $\alpha$. However for three SMC O-type supergiants Puls et al. (1996)
derived an exponent  of 0.52, whilst for 6 O-type dwarfs
analysed by Bouret et al. (2003) a value of 0.6 was determined.
Unfortunately due to the dependence of mass-loss on the stellar
luminosity and mass, a comparison of the known Galactic B-type
supergiant mass-loss rates and those derived here will not depend
solely on their different metallicities.

\begin{figure}
\begin{center}
\epsfig{file=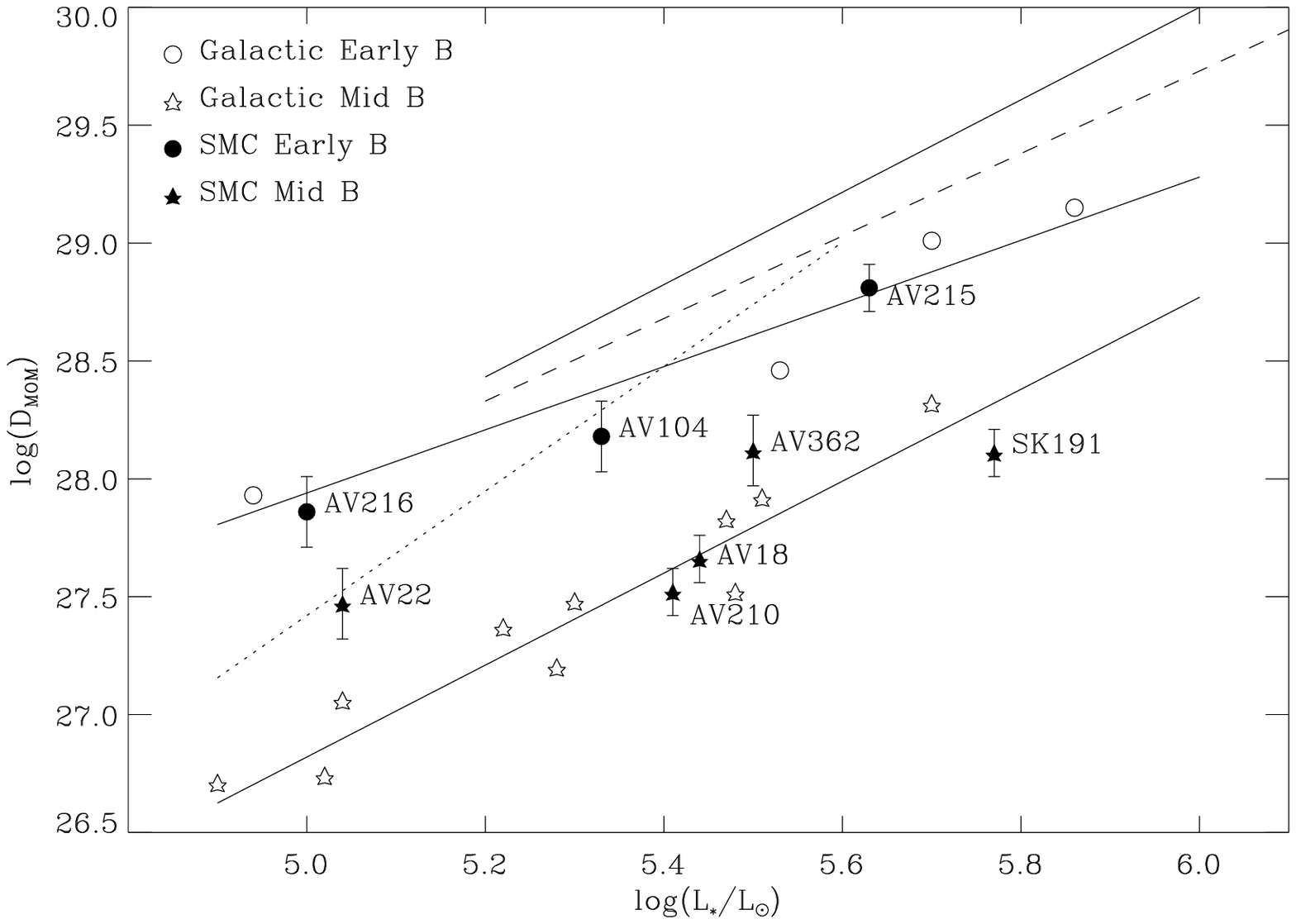, height=70mm, width=88mm, angle=0}
\caption[]
{
Wind momentum as a function of Luminosity. The solid lines are the
wind-momentum-luminosity relationship, in decreasing values of wind momentum,
from galactic O-type supergiants without clumping and with line-blanketing
(Repolust, Puls, \& Herrero 2003) and from galactic early and mid B-type
supergiants (Kudritzki \& Puls 2000). The dashed line (- - -) represents the
unblanketed WLR for O-type supergiants determined by Puls et al. (1996), 
whilst the dotted line ($\cdot\cdot\cdot$) represents that for the A-type
supergiants determined by Kudritzki et al. (1999). Note the galactic BA-type
WLR's were calibrated using an unblanketed version of {\sc fastwind}. The plot
shows the early and mid B-type stars of the galaxy ($\circ$, unfilled stars ;
Kudritzki \& Puls 2000) and the SMC ($\bullet$ , $\star$ ; this work). 
}
\label{wlr}
\end{center}
\end{figure}

\begin{figure*}
\begin{center}
\hspace{-2mm}
\epsfig{file=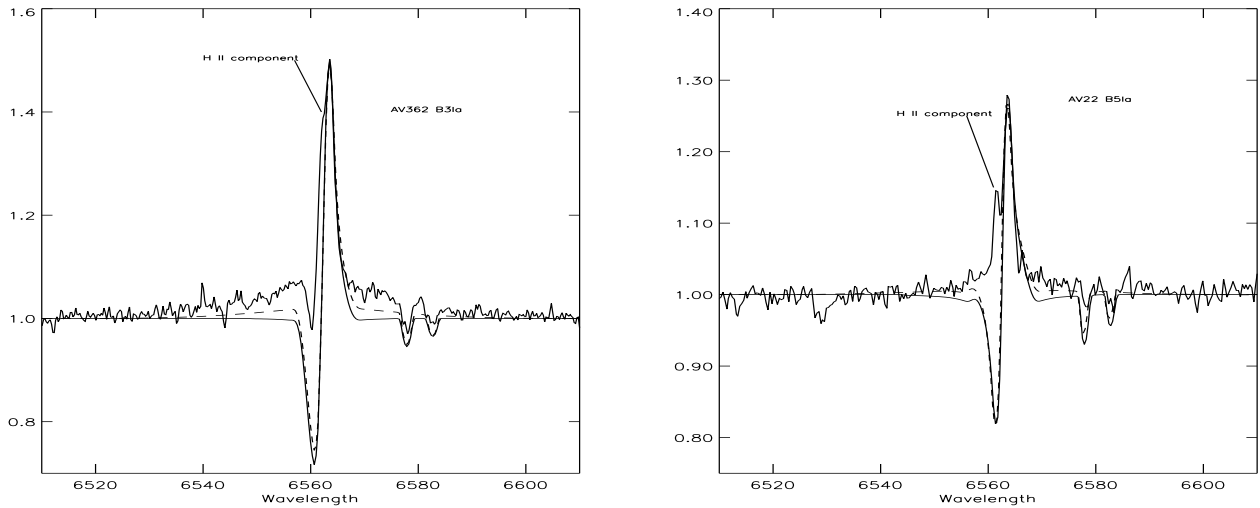, height=70mm, width=180mm, angle=0}
\caption[]
{
To show the effect of incoherent electron scattering on the wings of
AV362 and AV22. Shown are the best fits (solid line; -----) using the
parameters presented in Table~\ref{smcpar1} \& \ref{smcpar2} and the same model
with the inclusion of electron scattering (dashed line; - - -). Note the
improvement in the fit to the wings of AV22 by including electron scattering
and the little effect the scattering has on the fit to AV362. }
\label{ha2}
\end{center}
\end{figure*}

Vink et al. (2001) provide a recipe to determine the mass-loss rate which
includes a metallicity dependence of 0.85. Three different formulae are used to
describe the stellar mass-loss depending on its position relative to the so
called `bi-stability jumps', which vary due to the temperature and density
regime of the star. In Fig.~\ref{vink} we compare our observed mass-loss rates
with these theoretical predictions. Note that for the early B-type stars (B0 -
B1) Vink's predictions, which are based on the recipe for stars above the first
jump, underestimates \.{M} by a factor of 4, whilst for the mid B-type stars
the recipe for between the two bistability jumps overestimates \.{M} by a
factor of 10. This underestimation for the early B-type supergiants has also
been observed for O-type supergiants and dwarfs by Bouret et al. (2003) and
Evans et al. (2003). If we use the steeper exponent of 0.5 for the metallicity
dependence in Vink's recipes we get good agreement with the mass-loss rates for
the early B-type supergiants but the rates for the mid B-type supergiants are
now over estimated by a factor of 16. Clearly, a better understanding of the
metallicity and temperature dependence of mass-loss rates for B-type
supergiants is required.  

An alternative perspective is provided by considering the wind momentum,
D$_{\rm mom}$ = (\.{M}v$_{\infty}$R$^{1/2}$). As described in the introduction,
radiative line-driven wind theory relates the wind-momentum of a star to its
luminosity. In recent years massive star analyses have endeavoured to calibrate
this relationship as a function of spectral type and metallicity, with the plan
to invoke it as a method of deriving distances from OBA-type supergiants out to
the Virgo \& Fornax clusters (Puls et al. 1996; Kudritzki et al. 1999, 2000;
Crowther et al. 2002; Bouret et al. 2003; Bresolin et al. 2001; Urbaneja et al.
2003; Evans et al. 2003). In Fig.~\ref{wlr} we compare the wind-momenta of the
SMC B-type supergiants/giant studied here with the Galactic analysis of
Kudritzki et al. (1999). The solid lines in this figure represent the Galactic
WLR relationships as derived from the recent line-blanketed O-type star
analysis of Repolust, Puls, \& Herrero (2003) and the unblanketed early and mid
B-type supergiant analyses of Kudritzki et al. (1999, 2000). Surprisingly, the
SMC stars from B0 - B2 agree well with the galactic relationship. However the
two cool stars (B3 - B5) have wind-momenta which lie above the mid B
relationship, this is more significant in the case of the B5 star, AV22.  We
must note that the mid B WLR of Kudritzki et al. (1999) was based on spectral
types no later than B3, with the coolest star in their sample, HD 14134, being
at 18 kK in comparison to the 14 kK of AV362. From Fig.~\ref{wlr} we see that
AV22 agrees well with the WLR calibrated by Kudritzki et al. (1999) for A-type
stars. It is plausible that this cooler spectral type should follow the A-type
star WLR or indeed a different calibration of the WLR for late B-type stars.
Yet without more stars of similar or cooler spectral types (B5-B9), we can not
make any definite conclusions. Moreover the inclusion of clumping  into our
analysis would decrease the mass-loss rate in such a way as to bring the
wind-momenta of AV362 and AV22 into better agreement with the mid B WLR.

From Fig.~\ref{ha} it is apparent that the (far) wings of H$_{\alpha}$ in these
stars are quite strong and are unaccounted for by the {\sc fastwind} models. 
In Fig.~\ref{ha2} we show the results of including incoherent electron
scattering in the {\sc fastwind} calculations. The electron scattering only
increases the strength of the continuum around the H$_{\alpha}$ by 2\% and 1\%
in AV362 and AV22, respectively, where as the observed spectrum shows a raise
in the H$_{\alpha}$ wings of 7\% and 3\%. Whilst the situation has improved for
AV22, the electron scattering does little to improve the fit to H$_{\alpha}$ in
AV362. Note that the fact that the electron-scattering can not account for the
full strength of the electron scattering wings most probably is not related to
any presence or non-presence of wind-clumping, since these wings are formed in
the upper photosphere (cf. Kudritzki et al. 1999).

The agreement of our analysis with the galactic WLR for early B-type stars
maybe somewhat misleading as the analysis of Kudritzki et al. (1999) was
undertaken using an unblanketed version of {\sc fastwind}.  Repolust, Puls, \&
Herrero (2003) and Puls et al. (1996) have  carried out a line-blanketed and
unblanketed analysis, respectively, of a set of Galactic O stars. In
Fig.~\ref{wlr} the regressions fits of both analyses are plotted and imply that
the line-blanketed version of the O-type star WLR is steeper. Additionally for
a given luminosity, higher wind-momenta are required than for the unblanketed
results. Since a similar effect on the WLR is expected for the B-type stars, it
is therefore possible that this is masking any predicted metallicity effect. In
addition the uncertainty in the adopted distance to the SMC introduces a
systematic error  to the derived radii and mass-loss rates, this will also
effect the wind-momentum luminosity relationship. This changes the WLR through
the wind-momentum and the luminosity, shifting the relationship along both the
horizontal and vertical directions of Fig.~\ref{wlr}. 

In general, the fitting uncertainties  on the mass-loss rates are of the order
10 - 15 \% , however the assumption of no clumping in our models introduces an
additional uncertainty. Crowther et al. (2002) and Evans et al. (2003) noticed
some indication of clumping in their O- and B-type supergiants from
observations in the UV and far UV (see also Massa et al. 2003, Hillier et al.
2003). Moreover Repolust, Puls \& Herrero (2003) suggested the presence of
clumping to explain certain inconsistencies between observed and theoretically
predicted WLR's. Since the mass-loss rate varies with the filling factor as
\.{M}/$\surd$f~, assuming a filling factor of 0.1 reduces the mass-loss to
one-third of its value before considering clumping. Although clumping is an
important factor to be considered in the analysis of supergiants with thick
winds, the degree of clumping is still somewhat uncertain. Whilst for WR stars
the clumping enhances the strength of the coherent electron scattered wings of
the lines, this effect (which must not be confused with the incoherent electron
scattering wings discussed above) is too small for detection in the OB-type
supergiants (see Hillier et al. 2003). In addition by including clumping into
the analyses of massive stars the mass-loss rate will be reduced and hence we
would expect a shift in the WLR to lower wind-momenta (Repolust, Puls, \&
Herrero  2003).

\section{Conclusions} %

From high-resolution {\sc uves} data we have derived the atmospheric
and wind parameters of seven SMC B-type supergiants and one
giant in the spectral range B0 - B5. In addition we have carried
out an abundance analysis on each of these stars. This analysis
was undertaken using the non-LTE line-blanketed atmospheric
code {\sc fastwind}, which through spherical symmetry and
extended atmospheres allows for the computation of the stellar
wind (Santoloya-Rey et al. 1997; Herrero et al. 2002; Repolust,
Puls, \& Herrero 2003). 

The absolute abundances of the $\alpha$-processed elements (O, Mg \&
Si) and carbon in these SMC stars are in good agreement with SMC H
{\sc ii} regions and an unprocessed B-type star, AV304 (Kurt et al.
1999; Rolleston et al. 2003). The exception is nitrogen which in the
giant, AV216, is enhanced by a factor of four while there is a mean
overabundance of a factor of 13 present in the supergiants. Slightly
lower nitrogen enhancements are observed in B-type giants in NGC330
and A-type supergiants (Lennon et al. 2003; Venn et al. 1999).

A comparison was carried out with the most recent stellar evolution
models of Maeder \& Meynet (2001) which include rotationally induced
mixing. However we recalibrated these models to an SMC initial
nitrogen abundance of 6.55 dex as observed in AV304 instead of the
one-fifth solar implemented by Maeder \& Meynet. These models can
for the most part reproduce the observed nitrogen abundances if an
initial rotational velocity of 300 kms$^{-1}$ is invoked.
Nevertheless, observed projected rotational velocities are $<$ 100
kms$^{-1}$ and therefore we conclude that these models need to have
more efficient mixing at lower initial rotational velocities. In
addition we note that the evolutionary masses are generally higher
than the derived spectroscopic masses by at least a factor of 1.2.
Some of this discrepancy may be accounted for in the errors of the
adopted distance modulus and possibly the gravity, leaving a residual
difference in masses of less than 0.1 dex where M$_{evol}$ $>$ M$_{spec}$.

The theoretical mass-loss rate predictions of Vink et al. (2001)
which implement a metallicity dependence of Z$^{0.64}$,
underestimate \.{M} for the early B-type stars and overestimate it
for the mid B-types. By using a steeper metallicity dependence
(Z$^{0.5}$), the early B-type stars are in agreement but the
situation is then worse for the later spectral types. Clearly the
metallicity dependence of mass-loss rates for B-type stars is still
uncertain and a comparison of equivalent sets of stars in different
metallicity environments is needed to understand this dependence.

Finally, we have compared our results to the wind-momentum
relationships of Galactic B-type stars by Kudritzki et al. (1999,
2000). The wind momenta of our stars are in good agreement with the
galactic B-type supergiants. This is a puzzling result given the
theoretical predictions of a metallicity dependence on the
wind-momenta and the observational evidence of this effect for
O-type supergiants in the Magellanic Clouds (Puls et al. 1996;
Bouret et al. 2003). However the observed Galactic WLR was derived
using an unblanketed analysis which may mask any predicted
metallicity dependence when compared to our results. 

Work is currently underway to analyse EMMI spectra of additional SMC early
B-type supergiants in the spectral range B0 - B2.5. This larger sample of stars
will hopefully provide more insight into the behaviour of the wind-momenta and
stellar evolution of these early B-type supergiants at low metallicities.


\section*{Acknowledgements}   

CT is grateful to the Department of Higher and Further Education,
Training and Employment for Northern Ireland (DEFHTE) and the
Dunville Scholarships fund for their financial support. DJL
acknowledges funding from the UK Particle Physics and Astronomy
Research Council ({\sc pparc} under the grant PPA/G/S/2001/00131. We
would like to thank Robert Ryans for his continuous support with 
{\sc tlusty}. Based on observations collected at the European
Southern Observatory, Chile (Paranal).%

%


\begin{thebibliography}{}

\bibitem[1978]{Abb78}
Abbott, D.C. 1978, ApJ, 225, 893

\bibitem[1982]{Azz82}
Azzopardi, M.,\& Vigneau, J. 1982, A\&AS, 50, 291

\bibitem[1988]{Bec88}
Becker, S.R., \& Butler, K. 1988, A\&A, 201, 232

\bibitem[1989]{Bec89}
Becker, S.R., \& Butler, K. 1989, A\&A, 209, 244

\bibitem[1990]{Bec90}
Becker, S.R., \& Butler, K. 1990, A\&A, 235, 326

\bibitem[2003]{Bou03}
Bouret, J.-C., Lanz, T., Hillier, D.J., Heap, S.R., Hubeny, I.,
Lennon, D.J., Smith, L.J., \& Evans, C.J. 2003, ApJ,submitted

\bibitem[2001]{Bre01}
Bresolin, F., Kudritzki, R.-P., Mendez, R.H., \& Przybilla, N.
2001, ApJ, 548, L159

\bibitem[2002a]{Bre02a}
Bresolin, F., Gieren, W., Kudritzki, R.-P., Pietrzynski, G., \& Przybilla, N.
2002a, ApJ, 567, L277

\bibitem[2002b]{Bre02b}
Bresolin, F., Kudritzki, R.-P., Najarro, F.,Gieren, W., \& Pietrzynski, G., 
2002b, ApJ, 577, 107

\bibitem[1985]{But85}
Butler, K., \& Giddings, J.R. 1985, Coll. Comp. Project No. 7
(CCP7), Newsletter 9, London, p.7

\bibitem[1975]{Cas75}
Castor, J.I., Abbott, D.C., \& Klein, R.I. 1975, ApJ, 195, 157 

\bibitem[2002]{Cro02}
Crowther, P.A., Hillier, D.J., Evans, C.J., Fullerton, A.W., deMarco, O., \&
Willis, A.J. 2002, ApJ, 579, 774

 \bibitem[2000]{deM00}
de Mello, D.F., Leitherer, C., \& Heckman, T.M., 2000, ApJ, 530, 251

\bibitem[1994]{Den94}
Denissenkov, P. 1994, A\&A, 287, 113

\bibitem[2000a]{Duf00}
Dufton, P.L., McErlean, N.D., Lennon, D.J., \& Ryans, R.S.I. 2000,
A\&A, 353, 311

\bibitem[1988]{Ebe88}
Eber, F, \& Butler, K. 1988, A\&A, 202, 153

\bibitem[2002a]{Egg02}
Eggenberger, P., Meynet, G., \& Maeder, A. 2002, A\&A , 386, 576

\bibitem[2003]{Eva03}
Evans, C.J., Crowther, P.A., Fullerton, A.W., \& Hillier, D.J.
2003, MNRAS, submitted

\bibitem[2004]{Eva04}
Evans, C.J., Lennon, D.J., Trundle, C., Heap, S.R., \& Lindler, D.J 2004,
ApJ, submitted

\bibitem[1993]{Fitz93}
Fitzpatrick, E.L., \& Bohannan, B. 1993, ApJ, 404, 734

\bibitem[1990]{Fitz90}
Fitzpatrick, E.L., \& Garmany, C.D. 1990, ApJ, 363, 119

\bibitem[1987]{Gar87}
Garmany, C.D., Conti, P.S., \& Massey, P. 1987, AJ, 93, 1070

\bibitem[1992a]{Gie92}
Gies, D.R., \& Lambert, D.L. 1992, ApJ, 387, 673

\bibitem[1992]{Gra92}
Gray, D.F. 1992, The observation and analysis of stellar
photospheres, Cambridge university press, 2nd ed.

\bibitem[1981]{Ham81}
Hamann, W.-R. 1981, A\&A, 93, 353

\bibitem[2003]{Har03}
Harries, T.J., Hilditch, R.W., \& Howarth, I.D. 2003, MNRAS, 339, 157  

\bibitem[1995]{Has95}
Haser, S.M., Lennon, D.J., Kudritzki, R.-P.,
Puls, J., Pauldrach, A.W.A., Bianchi, L., \& Hutchings, J.B.
1995, A\&A, 295, 136

\bibitem[1998]{Has98}
Haser, S.M., Pauldrach, A.W.A., Lennon, D.J., Kudritzki, R.-P.,
Lennon, M., Puls, J., \& Voels, S.A. 1998, A\&A, 330, 285

\bibitem[2000]{Heg00}
Heger, A., \& Langer, N. 2000, ApJ, 544, 1016

\bibitem[1992]{Her92}
Herrero, A., Kudritzki, R.-P., Vilchez, J.M., Kunze, D., Butler,
K., \& Haser, S. 1992, A\&A, 261, 209

\bibitem[2002]{Her02}
Herrero, A., Puls, J., \& Najarro, F. 2002, A\&A, 396, 949

\bibitem[2002]{Her02a}
Herrero, A., \& Lennon, D.J. 2002, IAU symposium No. 215, Stellar
rotation, in press

\bibitem[1998]{Hil98}
Hillier, D.J., \& Miller, D.L. 1998, ApJ, 496, 407

\bibitem[2003]{Hil03}
Hillier, D.J., Lanz, T., Heap, S.R., Hubeny, I., Smith, L.J.,
Evans, C.J., Lennon, D.J., \& Bouret, J.C. 2003, ApJ, 588, 1039 

\bibitem[1997]{How97}
Howarth, I.D., Siebert, K.W., Hussain, G.A.J., \& Prinja, R.A.
1997, MNRAS, 284, 265

\bibitem[2001]{How01}
Howarth, I.D, Smith, K.C. 2001, MNRAS, 327,353

\bibitem[1995]{Hub95}
Hubeny, I., \& Lanz, T. 1995, ApJ, 439, 875

\bibitem[1978]{Kam78}
Kamp, L.W. 1978, ApJS, 36, 143

\bibitem[1991]{Kil91}
Kilian, J., Becker, S.R., Gehren, T. \& Nissen, P.E. 1991, A\&A,
244, 419

\bibitem[1980]{Kud80}
Kudritzki, R.-P. 1980, A\&A, 85, 174

\bibitem[1987]{Kud87}
Kudritzki, R.-P., Pauldrach, A.W.A., \& Puls, J. 1987, A\&A, 173, 293

\bibitem[1989]{Kud89}
Kudritzki, R.-P., Pauldrach, A.W.A., Puls, J, \& Abbot, D.C. 1989, A\&A, 219, 205

\bibitem[1992]{Kud92}
Kudritzki, R.-P. 1992, A\&A, 266, 395

\bibitem[1995]{Kud95}
Kudritzki, R.-P., Lennon, D.J., \& Puls, J. 1995, in Science with the
VLT, eds. J.R. Walsh \& I.J. Danziger, (Heidelberg: Springer), 246

\bibitem[1999]{Kud99}
Kudritzki, R.-P., Puls, J., Lennon, D.J., Venn, K.A., Reetz, J.,
Najarro, F., McCarthy, J.K., \& Herrero, A. 1999, A\&A, 350, 970

\bibitem[2000]{Kud00}
Kudritzki, R.-P., \& Puls, J. 2000, ARA\&A, 38, 613

\bibitem[1999]{Kur99}
Kurt, C.M., Dufour, R.J., Garnett, D.R, Skillman, E.D., Mathis,
J.S., Peimbert, M., Torres-Peimbert, S., \& Ruiz, M.-T 1999, ApJ, 518, 246

\bibitem[1987]{Lam87}
Lamers, H.J.G.L.M., Cerruti-Sola, M., \& Perinotto, M. 1987, ApJ, 314, 726

\bibitem[1995]{Lam95}
Lamers, H.J.G.L.M., Snow, T.P., \& Lindholm, D.M. 1995, ApJ, 455, 269

\bibitem[1995]{Lan95}
Langer, N., \& Maeder, A. 1995, A\&A, 295, 685

\bibitem[1992]{Lei92}
Leitherer, C., Robert, C., \& Drissen, L. 1992, ApJ, 401, 596

\bibitem[2001]{Lei01}
Leitherer, C., Le\~{a}o, J.R.S., Heckman, T.M., Lennon, D.J.,
Pettini, M., \& Robert, C. 2001, ApJ, 550, 724

\bibitem[1986]{Len86}
Lennon, D.J., Brown, P.L.F., Dufton, P.L., \& Lynas-Gray, A.E.
1986, MNRAS, 222, 719

\bibitem[1991]{Len91}
Lennon, D.J., Kudritzki, R.-P., Becker, S.T., Butler, K., Eber, F.,
Groth, H.G., \& Kunze, D. 1991, A\&A, 252, 498

\bibitem[1996]{Len96}
Lennon, D.J., Dufton, P.L., Mazzali, P.A., Pasian, F., \& Marconi,
G. 1996, A\&A, 314, 243

\bibitem[1997b]{Len97}
Lennon, D.J. 1997, A\&A, 317, 87

\bibitem[2003a]{LEN03}
Lennon, D.J., Dufton, P.L., \& Crowley, C. 2003, A\&A, 398, 455
 
\bibitem[1991]{Lyu91}
Lyubimkov, L.S. 1991, in IAU Symposium No. 145, Evolution of
stars: the photospheric abundance connection, eds. G. Michaud \& A.V.
Tutukov, (Dordrecht: Kluwer), 125, 145

\bibitem[2000c]{Mae00}
Maeder, A., \& Meynet, G. 2000, A\&A, 361, 159

\bibitem[2001]{Mae01}
Maeder, A., \& Meynet, G. 2001, A\&A, 373, 555

\bibitem[2002]{Mar02}
Martins, F., Schaerer, D., \& Hillier, D.J. 2002, A\&A, 382, 999

\bibitem[2003]{Mas03}
Massa, D., Fullerton, A.W., Sonneborn, G, \& Hutchings, J.B 2003, ApJ, 586, 996

\bibitem[2002]{Mas02}
Massey, P. 2002, ApJS, 141, 81

\bibitem[1963]{Mat63}
Matthews, T.A., \& Sandage, A.R. 1963, ApJ, 138, 30

\bibitem[1998]{McE98}
McErlean, N.D., Lennon, D.J., \& Dufton, P.L. 1998, A\&A, 329, 613

\bibitem[1999]{McE99}
McErlean, N.D., Lennon, D.J., \& Dufton, P.L. 1999, A\&A, 349, 553

\bibitem[1972]{Mih72}
Mihalas, D. 1972, ApJ, 177, 115

\bibitem[1996]{Mon96}
Monteverde, M.I., Herrero, A., Lennon, D.J, \& Kudritzki, R.-P.
1996, A\&A, 312, 24

\bibitem[1997]{Mon97}
Monteverde, M.I., Herrero, A., Lennon, D.J, \& Kudritzki, R.-P.
1997, ApJ, 474L, 107

\bibitem[1998]{Mon98}
Monteverde, M.I., \& Herrero, A. 1998, Ap\&SS, 263, 171

\bibitem[2000]{Mon00}
Monteverde, M.I., Herrero, A., Lennon, D.J 2000, ApJ, 545, 813

\bibitem[1972]{Osm72}
Osmer, P.S. 1972, ApJ, 171, 393

\bibitem[1973]{Osm73}
Osmer, P.S. 1973, ApJ, 184, 127

\bibitem[1999b]{Pau99}
Pauldrach, A.W.A., Hoffmann, T.L., \& Lennon, M. 2001, A\&A, 375, 161

\bibitem[2000]{Pet00}
Pettini, M., Steidel, C.C., Adelberger, K.L., Dickinson, M., \&
Giavalisco, M. 2000, ApJ, 528, 96

\bibitem[1998]{Pri98}
Prinja, R.K., \& Massa, D.L. 1998, in ASP Conf. Ser.
131, Properties of Hot Luminous Stars,ed I. Howarth (San
Francisco:ASP), 218

\bibitem[1968]{Prz68}
Przybylski, A. 1968, MNRAS, 139, 313

\bibitem[1971]{Prz71}
Przybylski, A. 1971, MNRAS, 152, 197

\bibitem[1996]{Pul96}
Puls, J., et al. 1996, A\&A, 305, 171

\bibitem[2000]{Pul00}
Puls, J., Springmann, U., \& Lennon, M. 2000, A\&AS, 141, 23

\bibitem[2003]{Rep03}
Repolust, T., Puls, J., \& Herrero, A. 2003, A\&A, submitted

\bibitem[2003]{Rol03}
Rolleston, W.R.J., Venn, K., Tolstoy, E., \& Dufton, P.L. 2003,
A\&A, 400, 21

\bibitem[1990]{Ruc90}
Rucinski, S.M. 1990, PASP, 102, 306

\bibitem[2002]{Rya90}
Ryans, R.S.I.R., Dufton, P.L., Rolleston, W.R.J., Lennon, D.J.,
Keenan, F.P., Smoker, J.V., \& Lambert, D.L. 2002, MNRAS, 336, 577

\bibitem[1997]{San97}
Santolaya-Rey, A.E., Puls, J., \& Herrero, A. 1997, A\&A, 323, 488

\bibitem[1968]{Sad68}
Sanduleak, N. 1968, AJ, 73, 246

\bibitem[1992]{Sch92}
Schaller, G., Schaerer, D., Meynet, G., \& Maeder, A. 1992, A\&AS, 96, 269
 
\bibitem[2001]{Sma01}
Smartt, S.J., Crowther, P.A., Dufton, P.L., Lennon, D.J., Kudritzki, R.-P.,
Herrero., A., McCarthy, J.K., \& Bresolin, F. 2001, MNRAS, 325, 257

\bibitem[1997]{Tal97}
Talon, S., Zahn, J.P., Maeder, A., \& Meynet, G. 1997, A\&A, 322, 209

\bibitem[2002]{Tru02}
Trundle, C., Dufton, P.L., Lennon, D.J., Smartt, S.J., \& Urbaneja, M.A.
2002, A\&A, 395, 519

\bibitem[2003]{Urb03}
Urbaneja, M.A., Herrero, A., Bresolin, F., Kudritzki, R.-P.,
Gieren, W., \& Puls, J. 2003, ApJ, 584, L73

\bibitem[2003]{Vaz03}
Vazquez, G.A., Leitherer, C., Heckman, T.M., Lennon, D.J., deMello,
D., Meuerer, G., \& Martin, C. 2003, ApJ, submitted

\bibitem[1995]{Ven95}
Venn, K.A. 1995, ApJ, 99, 659

\bibitem[1999]{Ven99}
Venn, K.A. 1999, ApJ, 518, 405

\bibitem[2000]{Ven00}
Venn, K.A., McCarthy, J.K., Lennon, D.J., Przybilla, N., Kudritzki,
R.-P., \& Lemke, M. 2000, ApJ, 541, 610

\bibitem[2001]{Ven01}
Venn, K.A.,et al. 2001, ApJ, 547, 765

\bibitem[2003a]{Ven03a}
Venn, K.A., Tolstoy, E., Kaufer, A., Skillman, E.D, Clarkson, S.M., Smartt, S.J.,
Lennon, D.J., \& Kudritzki, R.-P. 2003a, AJ, 126, 1326s

\bibitem[2003]{Ven03b}
Venn, K.A., \& Przybilla, N. 2003b, in ASP Conf. Ser. 304, CNO in the
Universe, eds. Charbonnelm C., Schaerer, D., Meynet, G., (ASP, San
Francisco), in press (astro-ph/0212258)

\bibitem[2001]{Vin01}
Vink, J.S., deKoter, A., \& Lamers, H.J.G.L.M. 2001, A\&A, 369, 574

\bibitem[1989]{Voe89}
Voels, S.A., Bohannan, B., Abbott, D.C., \& Hummer, D.G. 1989, ApJ,
340, 1073

\bibitem[1998]{Vra98}
Vrancken, M. 1998, PhD. thesis, Vrije Universiteit Brussel

\bibitem[2000]{Vra00}
Vrancken, M., Lennon, D.J., Dufton, P.L., \& Lambert, D.L. 2000,
A\&A, 358, 639

\bibitem[1995]{Wal95}
Walborn, N.R., Lennon, D.J., Haser, S.M., Kudritzki, R.-P., \&
Voels, S.A. 1995, PASP, 107, 104

\bibitem[2000]{Wal00}
Walborn, N.R., Lennon, D.J., Heap, S.R., Lindler, D.J., Smith,
L.J., Evans, C.J., \& Parker, J.W. 2000, PASP, 112, 1243

\bibitem[2001]{Wel01}
Wellstein, S., Langer, N., \& Braun, H. 2001, A\&A, 369, 939

\bibitem[1972]{Wol72}
Wolf, B. 1972, A\&A, 20, 275

\bibitem[1973]{Wol73}
Wolf, B. 1973, A\&A, 28, 335

\end{thebibliography}
\end{document}